\pdfoutput=1
\documentclass[acmsmall]{acmart}

\usepackage{algorithm}
\usepackage{algorithmic}
\usepackage{graphicx}
\usepackage{subcaption}
\usepackage{textcomp}
\usepackage{multirow}
\usepackage[normalem]{ulem}
\usepackage{xcolor}
\usepackage{booktabs}
\usepackage{tabularx}
\usepackage{colortbl}
\usepackage{array}
\usepackage{enumitem}
\usepackage{xspace}
\usepackage{amsmath}
\usepackage{makecell}
\usepackage{tcolorbox}

\newcommand{\ourTool}{MoEVD\xspace}
\newcommand{\ourToolBert}{MoEVD$_{CodeBERT}$\xspace}

\newcommand{\ourToolUniXcoder}{MoEVD$_{UniXcoder}$\xspace}

\newcommand{\CausalBert}{CausalVul$_{CodeBERT}$\xspace}

\newcommand{\CausalUniXcoder}{CausalVul$_{UniXcoder}$\xspace}

\newcommand{\rqone}{\textit{RQ1: How effective is \ourTool compared with SOTA approaches in vulnerability detection?}\xspace}

\newcommand{\rqtwo}{\textit{RQ2: How effective is the MoE framework within \ourTool?}\xspace}

\newcommand{\rqthree}{\textit{RQ3: How effective is \ourTool in detecting each CWE type and handling long-tailed distributions compared to SOTA approaches?}\xspace}

\newcommand{\rqfour}{\textit{RQ4: What is the impact of varying the number of experts on the performance of \ourTool?}\xspace}

\newcommand{\rqboxc}[1]{\begin{tcolorbox}[left=1pt,right=1pt,top=0pt,bottom=0pt,colback=gray!5,colframe=gray!40!black,before skip=5pt,after skip=0pt]#1\end{tcolorbox}}

\setcopyright{cc}
\setcctype{by}
\acmDOI{10.1145/3715736}
\acmYear{2025}
\acmJournal{PACMSE}
\acmVolume{2}
\acmNumber{FSE}
\acmArticle{FSE021}
\acmMonth{7}
\received{2024-09-11}
\received[accepted]{2025-01-14}

\begin{document}

\title{One-for-All Does Not Work! Enhancing Vulnerability Detection by Mixture-of-Experts (MoE)}

\author{Xu Yang}
\orcid{0000-0001-9963-6225}
\affiliation{%
  \institution{University of Manitoba}
  \city{Winnipeg}
  \country{Canada}
}
\email{yangx4@myumanitoba.ca}

\author{Shaowei Wang}
\orcid{0000-0003-3823-1771}
\affiliation{%
  \institution{University of Manitoba}
  \city{Winnipeg}
  \country{Canada}
}
\email{shaowei.wang@umanitoba.ca}

\author{Jiayuan Zhou}
\orcid{0000-0002-5181-3146}
\affiliation{%
  \institution{Huawei}
  \city{Toronto}
  \country{Canada}
}
\email{jiayuanzhou1@acm.org}

\author{Wenhan Zhu}
\orcid{0000-0001-6439-0720}
\affiliation{%
  \institution{Huawei}
  \city{Waterloo}
  \country{Canada}
}
\email{wenhanzhu1@acm.org}


\begin{abstract}
Deep Learning-based Vulnerability Detection (DLVD) techniques have garnered significant interest due to their ability to automatically learn vulnerability patterns from previously compromised code.
Despite the notable accuracy demonstrated by pioneering tools,
the broader application of DLVD methods in real-world scenarios is hindered by significant challenges.
A primary issue is the ``one-for-all'' design, where a single model is trained to handle all types of vulnerabilities.
This approach fails to capture the patterns of different vulnerability types, resulting in suboptimal performance,
particularly for less common vulnerabilities that are often underrepresented in training datasets.
To address these challenges, we propose \ourTool, which adopts the Mixture-of-Experts (MoE) framework for vulnerability detection.
\ourTool decomposes vulnerability detection into two tasks: CWE type classification and CWE-specific vulnerability detection.
By splitting the task, in vulnerability detection, \ourTool allows specific experts to handle distinct types of vulnerabilities instead of handling all vulnerabilities within one model.
Our results show that \ourTool achieves an F1-score of 0.44, significantly outperforming all studied state-of-the-art (SOTA) baselines by at least 12.8\%.
\ourTool excels across almost all CWE types, improving recall over the best SOTA baseline by 9\% to 77.8\%.
Notably, \ourTool does not sacrifice performance on long-tailed CWE types; instead,
its MoE design enhances performance (F1-score) on these by at least 7.3\%, addressing long-tailed issues effectively.

\end{abstract}

\ccsdesc[500]{Software and its engineering}

\keywords{Vulnerability Detection, Deep Learning, Mixture-of-Experts (MoE)}

\maketitle

\section{Introduction}\label{sec:introduction}

Vulnerability detection is a critical task in software engineering, aiming to identify security weaknesses that could be exploited by malicious entities. Deep learning-based vulnerability detection (DLVD) techniques have shown impressive results in academic settings~\cite{scandariato2014predicting,li2018vuldeepecker,chakraborty2021deep,hin2022linevd,li2021vulnerability,rahman2024towards,wen2024livable}. 

However, a recent study~\cite{wan2024bridging} reveals a significant challenge that prevents deep learning models from being adopted in industry settings. The challenge is the \textbf{one-for-all design limitation}. Existing DLVD techniques follow the ``one-for-all'' design, where a single model is trained to predict vulnerabilities across all types of code~\cite{fu2022linevul,li2018vuldeepecker,li2021sysevr,rahman2024towards,chakraborty2021deep,zhou2019devign,li2021vulnerability, hin2022linevd}. While this design simplifies the development and deployment process, it does not align with the diverse and specific needs of real-world applications. In practice, large organizations often encounter various types of vulnerabilities, each with unique characteristics and implications~\cite{wan2024bridging}. A one-for-all model may fail to adequately address the nuances of different vulnerability types and lack the capability of detecting certain types of vulnerabilities. 

\begin{figure}[!t]
    \centering
    \begin{minipage}[b]{0.5\columnwidth}
        \centering
    \includegraphics[bb=0 0 461 346, width=1.0\linewidth]{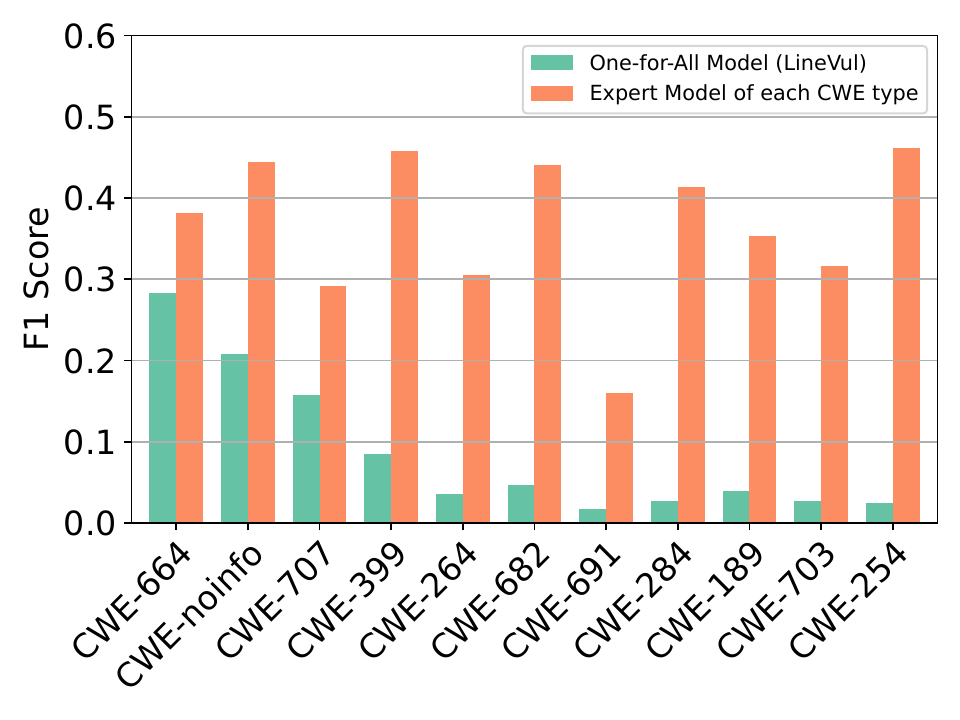}
    \caption{The performance of one-for-all Model (LineVul) vs. expert model trained on each specific CWE type across CWE types.}
    \label{fig:preliminary}
    \end{minipage}
    \hfill
    \begin{minipage}[b]{0.45\columnwidth}
        \includegraphics[bb=0 0 234 146, width=1.0\linewidth]{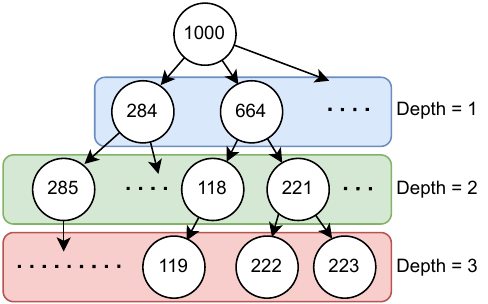}
	\caption{Hierarchy structure of CWE types. The arrows present the parent-child relationship between two CWE types. For instance, CWE-664 is the parent of CWE-221.}
	\label{fig:cwe-tree}
    \end{minipage}
\end{figure}

One significant issue stemming from the one-for-all design is the uneven and long-tailed distribution of different Common Weakness Enumeration (CWE)~\cite{CWE} types in real-world codebases. Certain types of vulnerabilities, such as those listed in the CWE Top 25~\cite{cwe.top.25}, are more prevalent and thus receive more attention in both research and industry. However, numerous other vulnerability types are less common and often underrepresented in training datasets. This imbalance leads to a situation where DLVD techniques are highly effective at detecting common vulnerabilities but struggle with rare ones. In practical terms, this means that while a model might excel in identifying frequent issues, its performance on less common, yet potentially critical vulnerabilities is subpar. This uneven performance can undermine the overall effectiveness of DLVD techniques, leaving some vulnerabilities undetected and increasing the risk of security breaches. For instance, Figure~\ref{fig:preliminary} presents the performance of LineVul~\cite{fu2022linevul}, a SOTA approach that is a single model trained to detect all types of vulnerabilities, on different CWE types compared with experts using the same model trained on each specific CWE type. As we can see, LineVul performs poorly on less frequent CWE types with F1-score ranging from 0.016 to 0.268. A similar trend can be observed on other one-for-all models as well. 


The Mixture-of-Experts (MoE) framework~\cite{jacobs1991textordfeminineadaptive,yuksel2012twenty} is a promising solution proposed by previous research to resolve the limitations of the one-for-all design.
The MoE framework addresses this limitation by first splitting the input space into sub-spaces, and then leveraging multiple specialized models (i.e., experts), trained to handle each sub-space. Instead of relying on a single model to handle the entire input space, MoE includes a router that dynamically assigns the most appropriate expert for handling a given input. 

To address the challenges mentioned above, we propose \ourTool, which leverages the Mixture-of-Experts (MoE) framework for vulnerability detection. Two key challenges need to be resolved to effectively leverage MoE in this context. First, how to split the input space effectively, so that each expert could focus on one type or a set of related vulnerabilities. Second, how to design a router that can effectively assign the input to the appropriate experts for handling the vulnerability detection task. To address the first challenge, we split the input code according to Common Weakness Enumeration (CWE) types~\cite{CWE}, a widely used classification system for software vulnerabilities. This splitting mechanism allows each expert to focus on learning patterns and detecting vulnerabilities specific to a particular type of CWE or a subset of CWE types. More specifically, to construct the expert for each CWE type, we fine-tune a pre-trained model (e.g., CodeBERT~\cite{feng2020codebert}) to predict if a piece of code is vulnerable. To address the second challenge, 
we train a router to dynamically select the appropriate experts for a given input code by formulating this task as a CWE type multi-class
classification task. Essentially, \ourTool decomposes the task of binary vulnerability detection into a CWE type classification task and a CWE-specific vulnerability detection task so that specific experts can be trained to handle a specific set of vulnerabilities instead of handling all types of vulnerabilities within one model.

We performed an extensive evaluation on a widely used benchmark dataset BigVul~\cite{fan2020ac}. Our results show that \ourTool achieves an F1-score of 0.44, significantly outperforming all studied state-of-the-art (SOTA) baselines by at least 12.8\%.
\ourTool outperforms the best SOTA baseline across almost all CWE types in terms of F1-score and recall. For instance, \ourTool improves the recall of the best SOTA baseline by a range from 9\% to 77.8\%. More importantly, \ourTool does not sacrifice performance in long-tailed CWE types. On the contrary, benefiting from the MoE design that enables the model to learn various vulnerability patterns individually, \ourTool improves the performance on vulnerabilities of long-tailed CWE types. For instance, \ourTool mitigates long-tailed issues by improving the F1-score of SOTA baselines on the long-tailed CWEs by at least 7.3\%. 

Our contributions are summarized below:
\begin{itemize}
    \item We are the first to adapt the Mixture-of-Experts (MoE) framework to enhance vulnerability detection and our extensive evaluation demonstrates its effectiveness, showing improvements across all CWE types.
    \item \ourTool enhances the detection of rare CWE types of vulnerabilities that are overlooked by existing one-for-all models and makes it more suitable for real-world applications. 
    \item To facilitate future research, we have made all the datasets, results, and code used in this study openly available in our replication package~\cite{MoEreplication}.
\end{itemize}

\section{Background and Motivation}\label{sec:background}
In this section, we introduce the background of our study, focusing on the pre-trained models for code representation Learning, CWE hierarchy, and Mixture-of-Experts (MoE) framework.


\subsection{Pre-trained Model for Code Representation Learning}
Deep learning-based vulnerability detection techniques rely on code representation learning to capture the syntactic and semantic properties of the code~\cite{yang2023does,zhou2019devign,li2021vulnerability}. These techniques transform code snippets into vectors, which can be effectively processed by classification models. Recently, several pre-trained models tailored for programming languages have emerged and demonstrated promising performance in code-related tasks, such as CodeBERT~\cite{feng2020codebert}, GraphCodeBERT~\cite{guo2020graphcodebert}, and UniXcoder~\cite{guo2022unixcoder}. Those models are based on transformer architecture, which has proven effective in code-related tasks. For instance, CodeBERT is built upon the foundations of BERT~\cite{devlin2018bert} through further pre-training on natural language (NL) - programming language (PL) pairs. Its design and training methods enable CodeBERT to understand not only the relationship between natural language and code but also the semantics of the source code. These pre-trained models have demonstrated SOTA performance in vulnerability detection~\cite{fu2022linevul, nguyen2022regvd}. In this study, we use pre-trained models as the foundation for our experts and router by fine-tuning them to adapt to downstream tasks. Specifically, we add a binary classification head to a pre-trained model for vulnerability detection and a multi-class classification head for CWE type multi-class classification.

\subsection{Common Weakness Enumeration}
Common Weakness Enumeration(CWE) is a category system for weaknesses and vulnerabilities. CWE follows a hierarchical structure, as illustrated in Figure~\ref{fig:cwe-tree}, where each CWE type has ancestors and related successors~\cite{CWE}. For instance, CWE-664 (Improper Control of a Resource Through its Lifetime) has two children, CWE-118 (Incorrect Access of Indexable Resource ('Range Error')) and CWE-221 (Information Loss or Omission). Typically, the children CWE types under a parent CWE type are related and share similar vulnerability patterns. A parent CWE is a more general category encompassing its child CWE types. For vulnerabilities where the details are unknown or unspecified, the NVD assigns a placeholder value, "CWE-noinfo," when it is unable to categorize a CVE under a specific CWE entry.

\subsection{Mixture-of-Experts (MoE)}\label{sec:MoE}

\begin{figure}[h]
\vspace{-0.1in}
	\includegraphics[bb=0 0 236 177, width=0.4\linewidth]{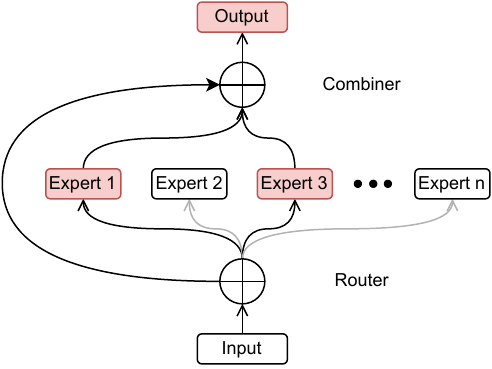}
	\caption{An Example Design of Mixture-of-Experts (MoE).} 
	\label{fig:MoE_design}
\vspace{-0.1in}
\end{figure}

\subsubsection{Mixture-of-Experts Framework Design}
Mixture-of-Experts (MoE) is a combinatory approach that leverages individual learning techniques as experts specializing in distinct sub-spaces of the input space~\cite{yuksel2012twenty}. MoE, proposed by Jacobs et al. in 1991~\cite{jacobs1991textordfeminineadaptive}, uses the idea of dividing the input space into sub-spaces, trains experts within each sub-space, and combines the knowledge of experts using a router, also called gating function. The core idea behind MoE is to select suitable experts for a particular input using a router. Typically, MoE contains the following three components:
\begin{itemize}
    \item \textbf{Experts:} These are individual models or neural networks trained to specialize in different parts of the input space or in different tasks. Experts can be of various types, such as linear models, decision trees, or deep neural networks.
    \item \textbf{Router (Gating function):} The router is a model that determines which expert or combination of experts should handle a particular input. It is typically implemented as a neural network and can be trained alongside the experts. Usually, not all experts are employed for inference; the router selects the most appropriate subset of experts for each input.
    \item \textbf{Combiner:} The combiner takes the outputs of the experts, weighted by the probabilities or weights assigned by the router, and produces the final output. 
\end{itemize}

Let $\theta$ = \{$\theta_g$, $\theta_e$\} represent the set of parameters of MoE, where $\theta_g$ represent the parameters for the router, and $\theta_e$ denotes the parameters of experts. Given an input vector $X$ and an output vector $Y$,  the final probability of $P$ can be formulated as:
\begin{align}
        P(Y|X, \theta) &= Aggregation_{i=1}^{N}(P(Y,i|X,\theta)) \\
                        &= Aggregation_{i=1}^{N}(g(i|X,\theta_g)P(Y|i,X,\theta_e^i))
\end{align}
, where $N$ donotes the number of experts, $g(i|X,\theta_g)$ denotes the probability of selecting the $i$-th expert to handle input data $X$, $P(Y|i,X,\theta_e^i)$ is the probability of $Y$ generated by the $i$-th expert given $X$, and $\theta_e^i$ is the parameters for $i$-th expert. The function $Aggregation$ aggregates results from each expert. The training objective for the MoE framework is to train both $\theta_g$ and $\theta_e$\, or only $\theta_g$. 

Figure~\ref{fig:MoE_design} presents a typical design of MoE. The input is first fed to the router and the router decides the most appropriate experts to handle the input. In this case, Expert 1 and Expert 3 are selected. The results returned by those two experts are aggregated in the combiner and output as the final results. 

\subsubsection{Vulnerability Detection with Mixture-of-Experts}
The motivation for our approach stems from the significant limitations of the prevalent ``one-for-all'' design in existing deep learning-based vulnerability detection techniques. Those techniques train a single model to handle all types of vulnerabilities, which often fails to capture the unique patterns and characteristics of different vulnerability types. This is particularly problematic for rare vulnerabilities that are underrepresented in training datasets, leading to low performance and increased security risk.

In contrast, the MoE framework allows for specialization, where each expert focuses on detecting vulnerabilities specific to a particular CWE type. This inherent partitioning aligns well with the MoE approach, where each expert can be dedicated to a specific CWE type. By leveraging the MoE framework, our goal is to improve the overall accuracy and efficiency of vulnerability detection, making it more suitable for real-world applications where diverse and specific vulnerabilities need to be addressed effectively.

\section{Methodology}\label{sec:methodology}

\begin{figure*}[tbh!]
	\includegraphics[bb=0 0 1082 517, width=1.0\linewidth]{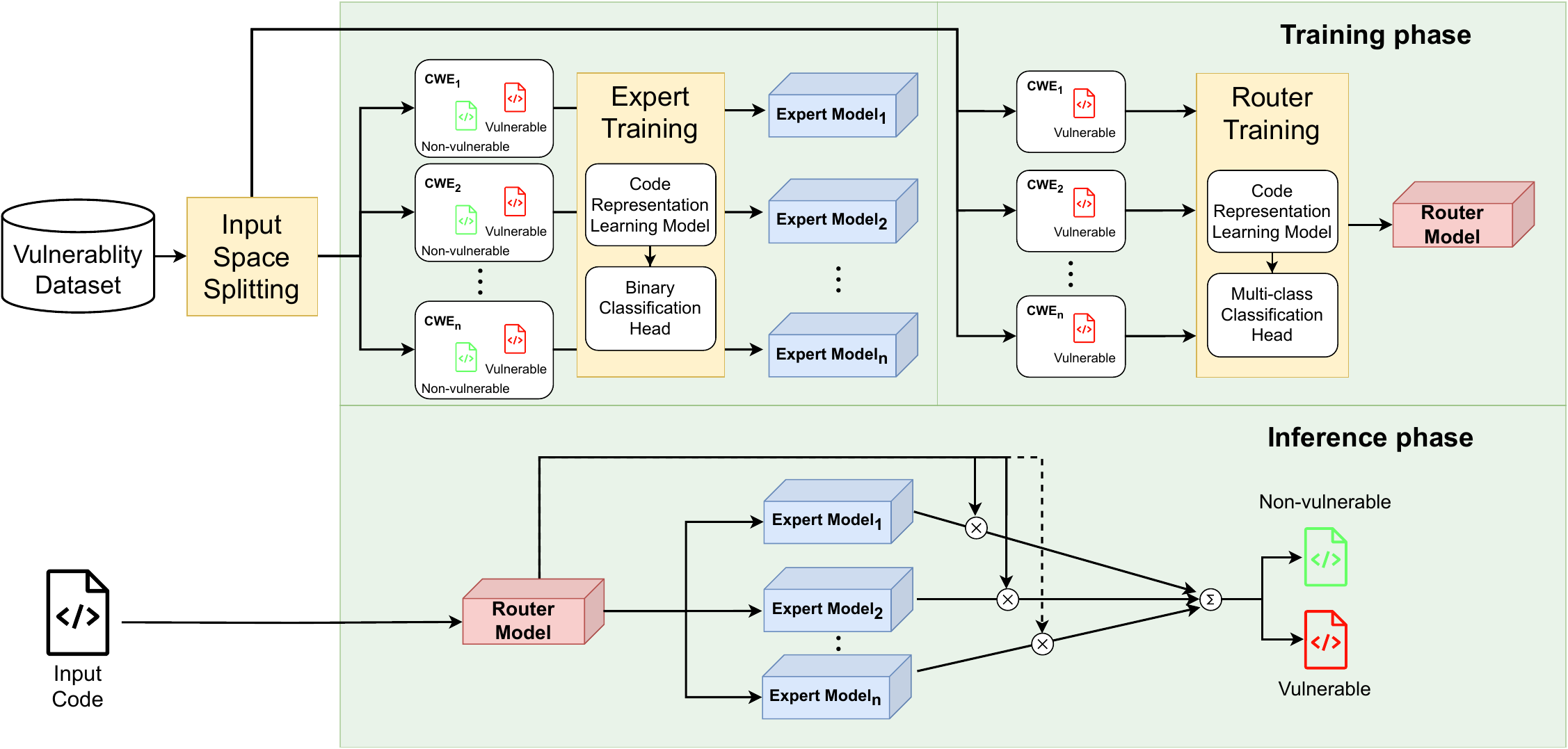}
	\caption{The pipeline of \ourTool.}
	\label{fig:pipeline}
\end{figure*}

In this section, we present the design and implementation of \ourTool. Figure~\ref{fig:pipeline} illustrates the overall pipeline of \ourTool, which leverages the Mixture-of-Experts (MoE) framework to identify and detect vulnerabilities with diverse patterns. During the training phase, we focus on training two key components within the MoE framework: the experts and the router. Each expert is specialized and trained to handle a specific CWE type (or combination) of vulnerabilities. Meanwhile, the router is trained using a CWE type classification task, enabling it to recognize the characteristics of specific CWE types and direct the input code to the appropriate experts. At inference time, the router assigns the input code to the most appropriate experts based on its characteristics. The selected experts then process the input code to predict vulnerabilities, and their outputs are aggregated to produce the final prediction. Essentially, \ourTool decomposes the task of binary vulnerability detection task into a CWE type classification task and a CWE-specific vulnerability detection task so that specific experts can be trained to handle a specific set of vulnerabilities instead of handling all types of vulnerabilities within one model.

\subsection{Splitting Input Space Based on CWE Types}\label{sec:CWEtypeConstruction}
The Mixture-of-Experts (MoE) framework provides a promising solution to the limitations of one-in-all models. To effectively leverage the MoE framework, it is crucial to split the input space into sub-spaces, allowing each expert to handle a specific sub-space. This division enables the specialization of experts, where each expert is trained to focus on and master the nuances of its designated sub-space, leading to improved performance in their respective areas. 

A common classification system to classify vulnerabilities is Common Weakness Enumeration (CWE) types, each representing a distinct category of software weaknesses, ranging from buffer overflows and injection flaws to improper authentication and information leakage~\cite{CWE}. These vulnerabilities of different CWE types exhibit different and often complex patterns that require specialized knowledge to detect effectively~\cite{aota2020automation, fu2023vulexplainer}.

To leverage the MoE framework, we split the input space of vulnerabilities into sub-spaces based on their CWE types. However, it is infeasible to train an expert for each specific CWE type directly for several reasons. First, there are too many CWE types (e.g., the BigVul dataset~\cite{fan2020ac} contains 88 CWE types). Training an expert for each individual CWE type would make the number of experts excessively large and incur significant overhead in expert training. Second, the distribution of different CWE types of vulnerabilities follows a long-tailed pattern, meaning some CWE vulnerability types are rare and have only a few vulnerable instances~\cite{zhou2023devil}.  It is challenging to train a robust expert on such rare CWE types.

To address the aforementioned challenges, we construct a CWE tree to present the hierarchical relationships and select the top-level CWE types as the basis for input space splitting. As introduced in Section\ref{sec:background}, CWE follows a hierarchical structure that naturally clusters CWEs with similar vulnerability patterns, and experts trained on the parent CWE types are capable of handling their child CWE types as well.

In this study, we begin by selecting the top-level CWE types (e.g., depth=1) as the basis for input space splitting. However, we found that several CWE types (e.g., CWE-388 and CWE-320) do not have sufficient instances. For example, in the BigVul vulnerability detection dataset, these types have less than 100 instances. For these cases, we aggregate all such top-level CWE types into a single category, named CWE-agg. This aggregation ensures that each CWE type has a sufficient number of instances for effective training while reducing the overall model complexity. We ended up with 12 CWE types. In other words, we need to train one expert for each of those CWE types with 12 experts in total.


\subsection{Training Phase}

As introduced in Section~\ref{sec:background}, an MoE framework typically has three components: experts, router, and combiner. To reduce the training complexity, we train the experts and the router separately. Below, we elaborate on the process of training experts and the router, and aggregating the output of experts into a final prediction.

\subsubsection{Expert training}\label{sec:expertTraining}
Vulnerabilities of the same CWE type share similar characteristics.
Therefore, for each CWE type, we train a specific expert model. For a given CWE type $CWE_i$, our goal is to train an expert model (Expert Model$_i$) to predict the probability of an input code $c$ being vulnerable $vul$, i.e., $\hat{y} = P(vul|CWE_i,c,\theta_e^i)$, where $\theta_e^i$ are the parameters for Expert Model$_i$. Assume we have a training dataset \(\mathcal{D}_{CWE_i} = \{(c_j, y_j)\}_{j=1}^{N_{CWE_i}}\) for the CWE type \(CWE_i\), where \(c_j\) is an input code instance and \(y_j\) is the corresponding ground truth label (i.e., vulnerable or non-vulnerable).
The overall loss for the expert model of CWE type \(CWE_i\) over the entire dataset can be defined as the cross-entropy loss:

\begin{equation}
    \mathcal{L}_{CWE_i}(\mathcal{D}_{CWE_i}, \theta_e^i) = -\frac{1}{N_{CWE_i}} \sum_{j=1}^{N_{CWE_i}} y_j \log(\hat{y}_j) 
\end{equation}

To construct the training data \(\mathcal{D}_{CWE_i}\), we consider the vulnerable code of the specific CWE type (i.e., $CWE_i$) as positive labels, while treating all other code (including those labeled as other CWE types and all non-vulnerable code) as negative. This data construction strategy allows each expert to specialize in detecting vulnerabilities related to its specific CWE type, providing a more focused and accurate detection model for each type. This strategy is a key aspect of \ourTool, distinguishing it from other vulnerability detection models that classify code simply as vulnerable or non-vulnerable without considering the specific CWE type (see more discussion in Section~\ref{sec:dis_trainingdataExpert}).

We construct experts using pre-trained models (e.g., CodeBERT) as the backbone and fine-tune an individual expert for each CWE type by following previous studies~\cite{fu2022linevul,rahman2024towards,yang2023does}. We initialize each expert model with the pre-trained weights and add a binary classification head. Then we optimize the model parameters $\theta_e$ by minimizing the overall loss function \(\mathcal{L}_{CWE_i}(\mathcal{D}_{CWE_i}, \theta_e^i)\).

\subsubsection{Router training}\label{sec:routerTraining}

The goal of this step is to train a router that can distinguish the code characteristics of different CWE types, so that the appropriate experts can be selected during routing. We formulate the task of router training as a CWE type multi-class classification task. Given an input code \(c\), the task is to classify its CWE type, \(CWE_i \in \{CWE_1, CWE_2, \ldots, CWE_i, \ldots\}^N\), where \(N\) is the number of CWE types. The training objective is to learn a model to predict the probability of the code \(c\) being categorized as a specific CWE type \(CWE_i\), \(g_i(CWE_i|c,\theta_g)\). To make the router proficient in differentiating between the features of different CWE types and recognizing the nuances of various vulnerabilities, we do not include non-vulnerable code in the training dataset. Let \(\mathcal{D}_R = \{(c_j, t_j)\}_{j=1}^{N_R}\) be the training dataset for the router, where \(c_j\) is an input code instance and \(t_j\) is the corresponding CWE type label, $N_R$ is the size of the training data for the router.
The probability of code \(c_j\) being classified as the $i$th class \(CWE_i\) by the router model \(g\) is given by:
\[
\hat{t}_{ji} = g(CWE_i | c_j, \theta_g)
\]

Note that the CWE types suffer from long-tail issues~\cite{zhou2023devil}, meaning CWE types are highly imbalanced. To mitigate this issue, we use Focal Loss~\cite{lin2017focal}, defined as:

\begin{equation}
    \mathcal{L}_{\text{focal}}(p_t) = -\alpha_t (1 - p_t)^\gamma \log(p_t)
\end{equation}

where \(p_t\) is the predicted probability for the true class. \(\alpha_t\) is a balancing factor for class \(t\) which is used to balance the importance of different classes. \(\alpha_t\) can be set higher for minority classes to ensure they contribute more to the loss, helping the model pay more attention to underrepresented classes. \(\gamma\) is a focusing parameter to down-weight easy examples and focus on hard examples.

We adopt Focal Loss to our multi-class classification task and the overall focal loss for the router model over the entire dataset \(\mathcal{D}_R\) can be defined as:

\begin{equation}
    \mathcal{L}_{R}(\mathcal{D}_R, \theta_g) = -\frac{1}{N_R} \sum_{j=1}^{N_R} \sum_{i=1}^{N} \alpha_i (1 - \hat{t}_{ji})^\gamma t_{ji} \log(\hat{t}_{ji})
\end{equation}

where \(N\) is the number of CWE types, \(\alpha_i\) is the balancing factor for class \(CWE_i\), \(\gamma\) is the focusing parameter, \(t_{ji}\) is the ground truth label for code \(c_j\) for class \(CWE_i\), and \(\hat{t}_{ji}\) is the predicted probability for code \(c_j\) for class \(CWE_i\).


Similar to expert training, we construct a router by fine-tuning a pre-trained model (e.g., CodeBERT) by adding a multi-class classification head. We select the pre-trained model because it has demonstrated proficiency in capturing the patterns from source code~\cite{yang2023does,zhou2019devign,li2021vulnerability}, which is essential for our router to assign input code to the appropriate experts based on their CWE patterns. We initialize the router model with the pre-trained weights and then optimize the parameters $\theta_g$ by minimizing the overall loss function.

\subsubsection{Combiner}
The combiner aggregates the output of the selected experts using a weighted summing mechanism:
\begin{align}
P(vul|c) = \sum_{i=1}^{K} Softmax(g(CWE_i|c,\theta_g))\ P(vul|CWE_i,c,\theta_e^i)
\end{align}
\(K\) is the number of experts that are selected to handle the code \(c\). In this study, we choose the top \(K\) experts most suitable for handling \(c\) based on the probabilities \(g_i(c, \theta_g)\) produced by the router. Instead of using all experts with different weights, using only the top \(K\) experts reduces the inference cost and reduces the impact/bias caused by irrelevant experts. Note that we use the probability of each expert produced by the router output as the weight to control the contribution from each selected expert. 

The \(Softmax()\) function ensures that the probabilities for the selected top \(K\) experts sum up to 1. For example, setting $K$ to 2 and the probabilities for the top 2 experts are 0.45 and 0.15. After Softmax normalization, their probabilities are 0.57 and 0.43, respectively. 

We investigate the impact of different $K$ values in \ourTool in Section~\ref{sec:rq4}. Our experiments suggest by selecting $K=2$, \ourTool provides the best trade-off between performance and inference time agreeing with existing practices on the selection of $K$~\cite{shazeer2017outrageously}.

\subsection{Inference Phase}

During the inference phase, the router processes both vulnerable and non-vulnerable input code. The router first assigns the input code to one or more experts based on the CWE classification result. Then the selected experts further predict whether the input code belongs to a specific vulnerability or not. The aggregation mechanism combines the outputs from the experts to make a final prediction on whether the code is vulnerable or non-vulnerable. By leveraging the specialized knowledge of each expert and the ability of the router to classify CWE types, \ourTool is able to outperform existing tools in vulnerability detection.

Note that our router is trained exclusively on vulnerable code, it may lack specific knowledge about non-vulnerable code. One question raised here is which expert will be assigned for the non-vulnerable code. According to the design of our router, non-vulnerable code will be assigned to the experts that handle vulnerable code with the most similar characteristics to that of the input non-vulnerable code, so that it can be further identified by the corresponding experts. However, even though the non-vulnerable code is assigned to other experts, we do not need to worry about this since we train all our experts including the data of the entire same set of non-vulnerable code. In theory, all experts should have consistent capability in identifying non-vulnerable code.

\section{Experiment Design}\label{sec:experiment}
In this section, we outline our research questions (RQs), describe the datasets used, define the evaluation metrics, present our analysis approach for each RQ, and provide implementation details.

\subsection{Research Questions}
We evaluate \ourTool across various dimensions to address the following research questions:
\begin{itemize}
    \item\rqone
    \hfill
    \item \rqtwo
    \hfill
    \item \rqthree
    \hfill
    \item \rqfour
\end{itemize}

In \textbf{RQ1}, we aim to assess the overall performance of \ourTool compared to current SOTA approaches. \textbf{RQ2} focuses on evaluating the effectiveness of the MoE framework in routing inputs to the most appropriate experts. \textbf{RQ3} examines \ourTool's effectiveness in detecting various CWE types and its ability to handle long-tailed distributions compared to SOTA approaches. \textbf{RQ4} investigates the impact of selecting different numbers of experts on the performance of \ourTool.

\subsection{DLVD Baselines}
In our study, we select two representative groups of DLVD baselines, graph-based models and transformer-based models:

\begin{itemize}
    \item \textbf{Graph-based models:} We first select three of the most commonly used models, SySeVR~\cite{li2021sysevr}, Devign~\cite{zhou2019devign} and Reveal~\cite{chakraborty2021deep}. SySeVR leverages the abstract syntax tree (AST) graph generated from code. The rest two models leverage graph neural network (GNN)~\cite{scarselli2008graph} to learn code feature representations and have shown promising results. We also employ LIVABLE~\cite{wen2024livable}, which is so far the state-of-the-art approach that combines graph-based model with text-based model.
    \item 
    \begin{sloppypar} \textbf{Transformer-based models:} We select two base models CodeBERT~\cite{feng2020codebert} and UniXcoder~\cite{guo2022unixcoder}, which have been shown to be effective in vulnerability detection~\cite{feng2020codebert,guo2022unixcoder,fu2022linevul}. Additionally, we select the recent SOTA approach CausalVul~\cite{rahman2024towards}, which introduced calculus-based causal learning. For a fair comparison, we implement CausalVul on these two models, resulting in \CausalBert and \CausalUniXcoder.
    \end{sloppypar}
\end{itemize}

\subsection{Experiment Dataset}
We select our experiment vulnerability detection dataset based on two criteria: 1) Diversity. They must be collected from a wide range of code repositories to maintain various types of vulnerability. 2) Annotations. They must contain CWE type or CVE-id annotation for training \ourTool. Therefore, we select the BigVul dataset~\cite{fan2020ac} as our experiment dataset, which is widely used in prior studies~\cite{yang2023does,li2021vulnerability,fu2022linevul,wen2024livable}. The BigVul dataset contains CVEs from 2002 to 2019, extracted from over 300 different open-source C/C++ projects, encompassing 88 different CWE vulnerability types. It contains 10,547 vulnerable functions and 168,752 non-vulnerable functions. Following the same experiment setting of previous studies~\cite{rahman2024towards,yang2023does}, we use the same 80\%/10\%/10\% split as train/val/test data for the BigVul dataset. We also use the same exact dataset partition used in CausalVul for a fair comparison. After aggregating CWE types as described in Section~\ref{sec:CWEtypeConstruction}, we obtain 12 CWE types.

\begin{table}[h]
\small
\caption{Overview of the studied dataset.}\label{tab:data}
\begin{tabular}{@{}lllrrrr@{}}
  \toprule
  \textbf{Dataset}       &\textbf{Granularity} & \textbf{\#Project} & \textbf{\#Vuln} & \textbf{\#Non-Vuln}  & \textbf{\#CWE} \\
  \midrule
  BigVul~\cite{fan2020ac}&function             & 348                & 10,547          & 168,752              & 88 \\
  \bottomrule
\end{tabular}

\end{table}


\subsection{Evaluation Metrics}

We use F1-score, precision, and recall as our evaluation metrics, which are widely used in previous studies~\cite{zhou2019devign,li2021vulnerability,chakraborty2021deep,rahman2024towards,yang2023does}. The F1-score provides a balance between precision and recall, offering a single metric that considers both false positives and false negatives. Precision measures the accuracy of positive predictions, while recall measures the ability to identify all relevant instances. Unlike previous research, we do not use accuracy as an evaluation metric due to the highly imbalanced nature of vulnerability detection datasets, which can result in misleading accuracy values dominated by the majority class~\cite{he2009learning}.

\subsection{Approach for Research Questions}
\subsubsection{Approach of RQ1}

\begin{sloppypar}
To demonstrate the effectiveness of \ourTool, we compare it with graph-based models (Devign~\cite{zhou2019devign}, Reveal~\cite{chakraborty2021deep}, LIVABLE~\cite{wen2024livable}) and transformer-based models (CodeBert~\cite{feng2020codebert}, UniXcoder~\cite{guo2022unixcoder}, and CausalVul~\cite{rahman2024towards}). Depending on different base transformer models (CodeBERT and UniXcoder), there are two variants of CausalVul, namely \CausalBert and \CausalUniXcoder. Similarly, we also implement \ourTool using two different base transformer models, resulting in two variants, namely \ourToolBert and \ourToolUniXcoder.
\end{sloppypar}

\subsubsection{Approach of RQ2}\label{sec:approach_rq2}

In RQ2, we evaluate the effectiveness of the MoE framework of \ourTool. To do that, we keep every trained expert untouched and compare \ourTool with two different variants: ensemble and random as described below. 

\begin{itemize}
    \item \textbf{Ensemble} Ensemble techniques train a machine learning (ML) model to fuse multiple models. We compare our MoE framework with ensemble techniques, replacing the MoE framework with ensemble techniques while keeping all other components unchanged. Specifically, we use the prediction probabilities of the positive class (vulnerable) from each expert model as input features for training an ensemble model. We select Random Forest (RF)\cite{biau2016random}, XGBoost\cite{chen2016xgboost}, CatBoost~\cite{prokhorenkova2018catboost}, and LightGBM~\cite{ke2017lightgbm} as baselines due to their widespread use and competitive performance in classification tasks~\cite{zhang2020learning, szczepanek2022daily, bentejac2021comparative}. For simplicity, we call those variants with different ensemble methods as Ensemble$_{RF}$, Ensemble$_{XGBoost}$, Ensemble$_{CatBoost}$, Ensemble$_{LightGBM}$.
    \item \textbf{Random Router:} To test the effectiveness of our router, we also compare \ourTool with a variant in which the router selects experts \textit{randomly}. In this variant, we replace our router with a random expert selector and keep all other components unchanged. We define this variant as \ourTool$_{random}$.

\end{itemize}

We evaluate the performance of the above two settings and compare them with \ourTool. In this RQ, we choose the best-performing implementation of \ourTool, \ourToolBert to compare with.

\subsubsection{Approach of RQ3} 
\begin{sloppypar}
To answer RQ3, we compare the performance between the selected best-performing baseline (\CausalBert) and \ourToolBert. We first analyze the performance (recall) of selected models in detecting vulnerabilities across each CWE type. BigVul dataset has 88 CWE types. Note that the model performance may not be stable enough on some CWE types which have an extremely small amount of vulnerabilities. In order to reduce the potential performance bias, we group the CWE types that have fewer than 10 vulnerable codes together as CWE-N$\leq\!10$. Note that for RQ3, CWE types are not aggregated by the CWE hierarchy; therefore, they are not the same as the CWE types used for training the expert models.
\end{sloppypar}


To evaluate the effectiveness of \ourTool in the long-tailed scenario, we first construct the long-tailed dataset. We split the BigVul dataset into two groups (the head group and the tail group) by following the settings from previous research~\cite{zhou2023devil}. More specifically, we categorize vulnerabilities into two groups: the head group, comprising the most frequent CWEs that constitute at least 50\% of BigVul, and the tail group, consisting of the remaining vulnerabilities. In the BigVul dataset, 4 CWE types (CWE-119, CWE-noinfo, CWE-20, and CWE-399) belong to the head group (56.2\% of vulnerabilities in total), while the other 84 CWE types belong to the tail group (43.8\% of vulnerabilities in total).
Finally, we compare the performance of \CausalBert and \ourToolBert between the two groups, respectively.



\subsubsection{Approach of RQ4}

In MoE, the number of experts $K$ is an important parameter.
In this RQ, we aim to investigate the impact of the value $K$ on \ourTool's performance.
Our goal is to find the value of $K$ that achieves the optimal balance between performance and computational cost.

\subsection{Implementation Details} \label{sec:exp_detail}
We downloaded the official pre-trained models and tokenizers for all evaluated transformer models from HuggingFace~\cite{wolf2019huggingface}. We implemented and conducted all experiments using PyTorch~\cite{paszke2019pytorch} and Autogluon~\cite{erickson2020autogluon}. To train the expert and router models, we used a batch size of 32, a learning rate of 1e-5, the AdamW~\cite{loshchilov2017decoupled} as the optimizer, and trained the model for 10 epochs. For Focal Loss, we use the default setting~\cite{focalloss}, i.e., setting the focusing parameter \(\gamma\) to 1 and the balancing factor \(\alpha_t\) as the inverse of its percentage of number of samples for each CWE type. To optimize computational efficiency, we used the bfloat16 (brain floating point) mixed precision. All experiments were conducted on a Linux server equipped with four Nvidia RTX 3090 GPUs, 24 CPU cores, and 128GB of memory. Each expert’s training time is approximately 1 hour, and the router model's training time is approximately 10 minutes. During inference, when we set the number of experts to $K=2$, \ourTool can scan more than 100 code snippets per second.
For training the ensemble model in\ref{sec:approach_rq2}, We used a state-of-the-art AutoML tool Autogluon~\cite{agtabular} for training the ensemble model. Autogluon trains ML models with hyperparameter tuning, bagging, stacking and other techniques to optimize the performance, we select the four highest-performing ensemble models as our baselines.

We reproduce all DLVD baselines by using the replication packages provided in their works~\cite{zhou2019devign,chakraborty2021deep,feng2020codebert,guo2022unixcoder,wen2024livable,rahman2024towards}. We also adopt the same hyperparameter settings in their works.



\section{Results} \label{sec:results}
\subsection{RQ1: \ourTool vs. DLVD Baselines}

\begin{sloppypar}
\textbf{\ourTool outperforms the best baseline by 12.8\%.}
Table~\ref{tab:rq1_baselines} presents the performance comparison between our approach and studied baselines. We observe that \ourTool outperforms all SOTA baselines, including graph-based and transformer-based models in all evaluated metrics (F1-score, precision, and recall). 
Specifically, \ourToolBert achieves the best F1-score of 0.44, improving baselines from 12.8\% to 69.2\%. When looking at recall, \ourToolBert also achieves the best performance (0.47), while in terms of precision, \ourToolUniXcoder achieves the best performance (0.47).

\end{sloppypar}


\begin{sloppypar}
\textbf{\ourTool is able to deliver the best performance for all base models.}
For both CodeBERT and UniXcoder variants of \ourTool, \ourTool outperforms the base model and the corresponding CausalVul variant.
\ourToolBert, which is the best-performing variant based on CodeBERT, achieves the highest F1-score of 0.44, achieving a 15.8\% improvement over CodeBERT with an F1-score of 0.38.
\ourTool also consistently outperforms other approaches in terms of recall, which indicates \ourTool is effective in recognizing more vulnerabilities, highlighting the effectiveness of \ourTool over existing one-for-all DLVD approaches.
\end{sloppypar}

\begin{table}[h]
\caption{The performance of different VD approaches on BigVul dataset in terms of studied metrics.}\label{tab:rq1_baselines}
\begin{tabular}{@{}llrrr@{}}
\toprule
\textbf{Model Type}                             & \textbf{Approaches}    & \textbf{F1-score}   & \textbf{Precision}     & \textbf{Recall} \\
\midrule
\multirow{4}{*}{Graph-based}                    & SySeVR                 & 0.19                &     0.31               &    0.14    \\
                                                & Devign                 & 0.27                &     0.30               &    0.25    \\
                                                & Reveal                 & 0.26                &     0.23               &    0.30    \\
                                                & LIVABLE                & 0.39                &     0.40               &    0.37    \\
\midrule
\multirow{6}{*}{\makecell[c]{Transformer-based}}& CodeBERT               &  0.38               &     0.45               &    0.33    \\
                                                & \CausalBert            &  0.39               &     0.43               &    0.36    \\
                                                & \ourToolBert           &  \textbf{0.44}      &  0.42                  &   \textbf{0.46}     \\
\cmidrule(rl){2-5}
                                                & UniXCoder              &   0.38              &     0.46               &   0.32     \\
                                                & \CausalUniXcoder       &   0.39              &     0.35               &        0.45\\
                                                & \ourToolUniXcoder      &  0.43               &   \textbf{0.47}        &   0.40     \\
\bottomrule
\end{tabular}%
\end{table}
\rqboxc{\ourTool significantly outperforms all SOTA baselines. \ourTool's implementation with the CodeBERT model (\ourToolBert) achieves the best F1-score of 0.44 with 12.8\% improvement over the best SOTA baseline. Additionally, \ourTool's implementation with the UniXcoder model (\ourToolUniXcoder) achieves the highest precision of 0.47 and a notable F1-score of 0.43.}
\subsection{RQ2: Effectiveness of MoE}


\begin{sloppypar}
\textbf{\ourTool outperforms variants with ensemble methods by at least 12.7\% F1-score.} Table~\ref{tab:rq2_ensemable} presents the comparison between \ourTool and its variants with ensemble methods. \ourTool significantly outperforms those variants in terms of F1-score and recall. Specifically, compared to the best-performing variant, Ensemble$_{XGBoost}$, \ourTool achieves a 12.7\% improvement in F1-score. These results suggest that the MoE framework is superior to other ensemble methods. A potential reason is that combining the results of all experts introduces additional noise, leading to less accurate final results. 
\end{sloppypar}

\begin{table}[]
\caption{The performance of variants with different ensemble methods and routers.}\label{tab:rq2_ensemable}

\begin{tabular}{@{}lrrr@{}}
\toprule
\textbf{Variant}        & \textbf{F1-score} & \textbf{Precision} & \textbf{Recall} \\
\midrule
Ensemble$_{LightBGM}$     & 0.38              & \textbf{0.54}      & 0.29 \\
Ensemble$_{CatBoost}$     & 0.39              & 0.50               & 0.31 \\
Ensemble$_{XGBoost}$      & 0.39              & 0.51               & 0.31 \\
Ensemble$_{RF}$           & 0.38              & 0.49               & 0.30 \\ 
\midrule
\ourTool$_{random}$       & 0.08              & 0.07               & 0.09 \\
\midrule
\ourTool                  &  \textbf{0.44}    &  0.42              &   \textbf{0.46}     \\
\bottomrule
\end{tabular}%
\end{table}


\begin{figure}[]
    \centering
    \includegraphics[bb=0 0 1008 720, width=0.8\linewidth]{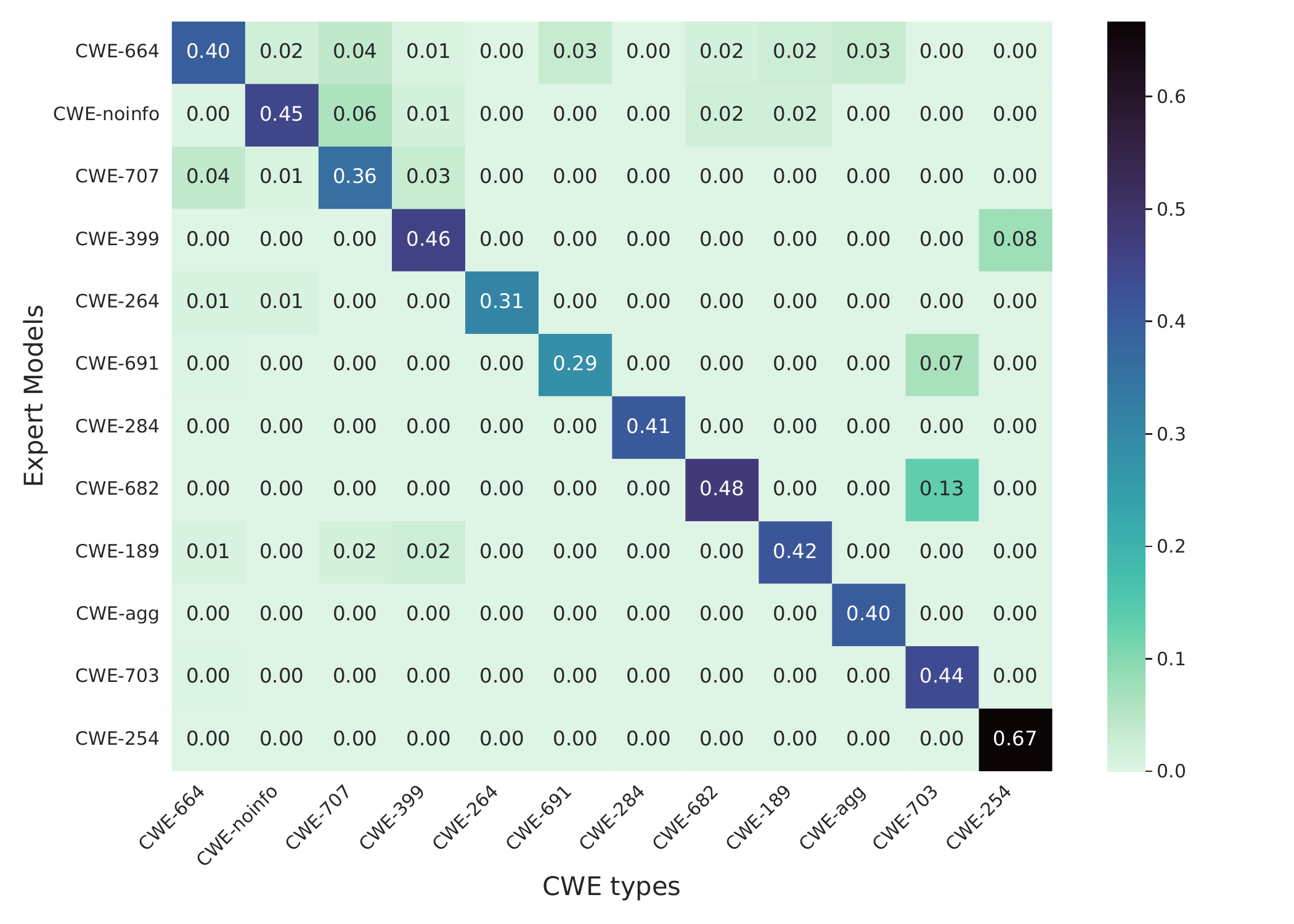}
    \caption{The performance matrix of each expert model on each CWE type in terms of F1-score. Darker colors indicate better performance.}
    \label{fig:heatmap}
\end{figure}

To understand the impact of the router, we compare the performance of \ourTool and \ourTool$_{random}$. From Table~\ref{tab:rq2_ensemable}, we observe that the performance of \ourTool$_{random}$ is extremely low (a F1-score of 0.08). Such a result indicates that selecting appropriate experts is crucial for our approach. 
We further measure the router's accuracy in selecting the correct expert for vulnerable input code. If the router can select the correct expert within the top K assigned experts, we consider it correct, otherwise, we consider it incorrect. The ground truth of an expert is determined by its CWE types after aggregation. If a vulnerable code of a CWE type A is assigned to an expert whose training data consists of CWE A, we consider the vulnerable code to be assigned correctly.
As shown in Table~\ref{tab:rq2_follow}, in 63.8\% of the cases, the router can select the correct experts. In other words, there is still room for improvement (see more details in Section~\ref{sec:potenialDirection}).


The router is an essential component of the MoE framework, correctly assigning the vulnerability to the correct CWE expert is important. To further understand how the router's performance affect the overall VD performance of \ourTool, we compare the performance between the cases where experts are correctly selected by the router and experts are wrongly selected. As the results shown in Table~\ref{tab:rq2_follow}, if the experts can be selected correctly, the performance is significantly better (by 133.7\%) than those who failed. 

The subpar performance when the expert is wrongly selected is expected due to the design of MoE. Each expert is trained on a specific CWE type and cannot identify the pattern of other CWE types. This can be observed in Figure~\ref{fig:heatmap}, where we show the performance of each expert across different CWE types. As expected, the experts only perform well on their own specific CWE type.


\begin{table}[]
\caption{The performance of \ourTool when the router can select experts correctly (Correctly selected) and wrongly (Wrongly selected).}
\label{tab:rq2_follow}
\begin{tabular}{@{}lrrrr@{}}
\toprule
\textbf{Expert Selection}  &  \textbf{\% vuln code}   & \textbf{F1-score} & \textbf{Precision} & \textbf{Recall} \\
\midrule
Correctly Selected         &  63.8\%                  & 0.44              & 0.36               & 0.56 \\
Wrongly Selected           &  36.2\%                  & 0.18              & 0.13               & 0.28  \\
\bottomrule
\end{tabular}%
\end{table}

\rqboxc{Mixture-of-Experts (MoE) design outperforms all ensemble models. The router plays a crucial role in \ourTool design, and significantly impacts overall \ourTool performance.}


\subsection{RQ3: Effectiveness of \ourTool across CWE Types}

\begin{figure}[!t]
    \centering
    \begin{minipage}[b]{0.49\columnwidth}
    \centering
    \includegraphics[bb=0 0 461 346,width=1\linewidth]{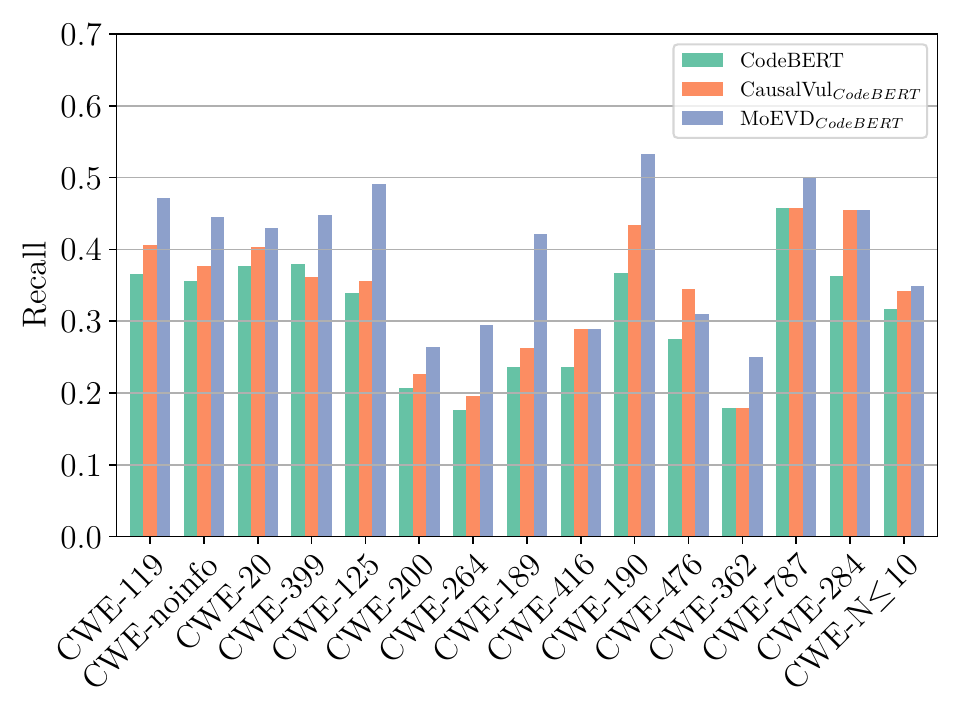}
    \caption{Recall of CodeBERT, \ourToolBert, and \CausalBert on different CWE types. The CWE types are displayed in descending order of frequency, from left to right.}
    \label{fig:rq3-fig}
    \end{minipage}
    \begin{minipage}[b]{0.49\columnwidth}
        \centering
    \includegraphics[bb=0 0 720 432, width=1\linewidth]{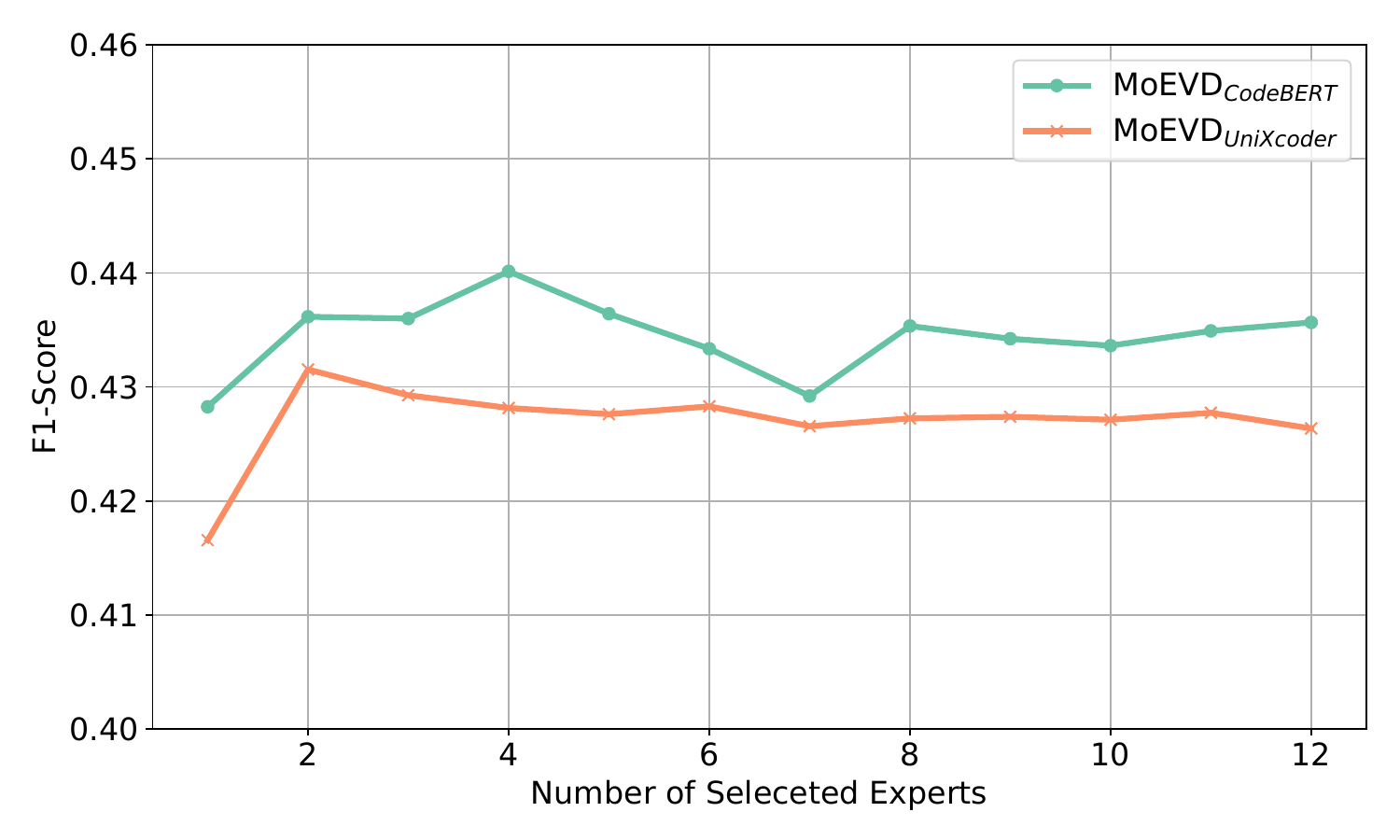}
    \caption{F1-score of \ourToolBert and \ourToolUniXcoder with different number of selected experts ($K$).}
    \label{fig:rq4-fig}
    \end{minipage}
\end{figure}




\begin{table*}[]
\caption{Performance achieved by CodeBERT, \ourToolBert, and \CausalBert on Head and Tail groups.}
\label{tab:rq3}
\begin{tabular}{llccccccc}
\toprule
                &  & \multicolumn{3}{c}{\textbf{Head}}      &                    & \multicolumn{3}{c}{\textbf{Tail}}      \\  \cmidrule(lr){3-5} \cmidrule(l){7-9}

                &  & \textbf{F1-score}             & \textbf{Precision} & \textbf{Recall} &  & \textbf{F1-score} & \textbf{Precision} & \textbf{Recall} \\ \midrule
CodeBERT        &  & 0.34                          & 0.30               & 0.38            &  & 0.27              & 0.24               & 0.32  \\
\CausalBert     &  & 0.34                          & 0.30               & 0.39            &  & 0.26              & 0.23               & 0.31  \\
\ourToolBert    &  & \textbf{0.37}                 & 0.29               & \textbf{0.50}   &  & \textbf{0.29}     & 0.23               & \textbf{0.42}  \\
\bottomrule
\end{tabular}%
\end{table*}

\begin{sloppypar}
\textbf{\ourTool outperforms baselines in most CWE types in terms of all studied metrics, typically recall and F1-score.}
Figure~\ref{fig:rq3-fig} presents the recall achieved by CodeBERT, \CausalBert, and \ourToolBert on all CWE types.
We observe that in most of the CWE types (14 out of 15), \ourToolBert outperforms CodeBERT and \CausalBert with a range of improvement from 9\% (CWE-787) to 77.78\% (CWE-189) in terms of recall. In terms of F1-score, we observe a similar trend that in 11 out of 15 CWE types, \ourToolBert outperforms \CausalBert, and 9 out of 15 in terms of precision. Due to the page limit, we put the complete results of F1-score and precision in our replication package~\cite{MoEreplication}.
\end{sloppypar}


\begin{sloppypar}
\textbf{\ourToolBert outperforms SOTA baselines \CausalBert and CodeBERT in both the head and the tail groups in terms of F1-score and recall, and achieves similar precision compared to SOTA baselines.}
Table~\ref{tab:rq3} presents the results of studied approaches on the head and the tail groups. We observe an improvement of \ourToolBert over the SOTA baselines in both the head and the tail groups. In terms of F1-score, \ourToolBert improves CodeBERT by 9.8\% and \CausalBert by 7.3\% in the head group. In the tail group, \ourToolBert improves CodeBERT by 7.3\% and \CausalBert by 11.4\%. When comparing recall, \ourToolBert achieves a significant improvement of 31.2\%, compared with baselines while maintaining a similar precision. The results indicate that, unlike baselines, when improving the binary classification performance, \ourTool does not sacrifice the performance in long-tailed CWE types. On the contrary, benefiting from the MoE design that enables the model to learn various vulnerability patterns individually, \ourTool improves the performance on the tail group.
\end{sloppypar}

\rqboxc{\ourToolBert outperforms SOTA baselines on most of the CWE types in terms of F1-score and recall. For instance, \ourToolBert improves the recall of SOTA baseline by a range from 9\% to 77.8\%. \ourToolBert also mitigates long-tailed issue by improving F1-score of SOTA baselines on the tail group at least by 7.3\%.}
\subsection{RQ4: Impact of the Number of Selected Experts}\label{sec:rq4}

\textbf{More experts do not yield significant performance improvement.}
Figure~\ref{fig:rq4-fig} presents the F1-score of \ourToolBert and \ourToolUniXcoder with a different number of selected experts ($K$). 
We observe that no significant advantages are achieved by increasing the number of experts.
The worst performance for both models occurs at $K=1$. \ourToolBert achieve its optimal performance when $K=4$, and \ourToolUniXcoder achieves its optimal performance at $K=2$. When $K$ is greater than 2, the increase of the number of selected experts does not yield significant benefits for both models. To balance the overhead and the performance, we set $K$ to 2 in this study. 

One possible reason for the lack of significant improvement with more experts is that each expert is only effective in its own CWE type and cannot identify the vulnerability for other types as indicated in Figure\ref{fig:heatmap}. Assigning additional experts will reduce the contribution of the most effective experts therefore adding more noise to the combiner.





\rqboxc{We do not observe significant advantages in increasing the number of experts. Employing a single expert yields sub-optimal results. Through our experiments, we find $K=2$ achieves an optimal balance between inference time and performance.}

\section{Discussion}
In this section, we discuss our special design in terms of expert training, the overhead, and the potential future research direction.
\subsection{Expert Training Task}
\label{sec:dis_trainingdataExpert}
Different from 1) classifying vulnerability types for a given vulnerability and 2) identifying whether a given input is a vulnerability or not, our experts learn vulnerability knowledge by identifying whether the given input belongs to a specific type of vulnerability or not.
As described in Section~\ref{sec:methodology}, when training an expert for a specific CWE type, we only consider the vulnerable code of that type as positive, and all other code (including other CWE type vulnerable code and non-vulnerable code) as negative.
By doing so, an expert is more focusing on learning the pattern of a specific type of vulnerability, improving the performance on distinguishing the target type of vulnerability.

\begin{sloppypar}
To validate our approach, we also trained a variant of \ourToolBert using only non-vulnerable data as the negative label, excluding code from other CWE types. The F1-score dropped from 0.44 to 0.41. This result indicates that using data from other CWE categories as negative helps the experts learn to distinguish between different types of vulnerabilities more effectively.
\end{sloppypar}

\subsection{Overhead of MoE Compared With One-for-All Model}
The training cost of \ourTool depends on the number of experts to train. For example, in the BigVul dataset, we have 12 CWE types and consequently we have 12 experts to train. Assume the total tunable parameters of a one-for-all model with the same base model (e.g. CodeBERT) is $N$. Then, \ourTool have a total of $12N + R$ tunable parameters, where $R$ is the tunable parameters of the router network. During inference, the computation cost depends on the number of experts $K$ as discussed in RQ4. Assume the inference cost of a one-for-all model is $x$, and we set $K=2$, then the total computation cost will be $2x + r$ where $r$ is the inference cost of the router. In our case, we use the same base model for experts and the router, and we set $K$ to 2. The inference cost is 3$x$. The time cost under our experiment setting can be found in section\ref{sec:exp_detail}.

\subsection{Potential Future Direction}\label{sec:potenialDirection}
Our research is the first study in vulnerability detection that employs the MoE framework. The design of \ourTool is straightforward, utilizing the vanilla MoE framework proposed by Jacobs et al.~\cite{jacobs1991textordfeminineadaptive}, which already demonstrates a significant improvement over the one-for-all design. In RQ2 (table~\ref{tab:rq2_follow}), we investigate the performance of the router, and we notice that only 63.8\% of the vulnerable code was correctly directed to the appropriate experts. In the ideal setting, if all the vulnerable code can be correctly assigned to the appropriate experts, the F1-score for \ourTool would be improved from 0.44 to 0.51. Thus, there remains a large margin to improve with the MoE framework and we believe that leveraging MoE in vulnerability detection is key to making DLVD methods more practical in real-world scenarios. One possible future direction is to improve the accuracy of the router to select the appropriate experts. Another possible direction is to leverage a more advanced MoE framework~\cite{shazeer2017outrageously,lepikhin2020gshard,fedus2022switch}. We encourage future research to investigate those two directions. 

\subsection{Threats to Validity}
\subsubsection{Internal validity}
\ourTool’s performance may be influenced by the distribution of CWE types within the training data, particularly for unseen CWEs. To mitigate this potential bias, we trained \ourTool on the BigVul dataset, which is one of the most comprehensive datasets, containing over 10,000 vulnerability instances across 88 distinct CWE types. For vulnerabilities belonging to unseen CWE types, \ourTool can be easily extended by training and adding specific experts for new vulnerabilities and re-training the router. 

\subsubsection{External validity}
Threats to external validity relate to the generalizability of our findings. In this study, we evaluate \ourTool using CodeBERT and UniXCoder as base models due to their widespread use in code-related tasks. While our findings might not be generalized to other models. However, As a framework, \ourTool is designed to be flexible and can integrate any model as an expert, future research is encouraged to explore the performance of \ourTool with a broader range of models.

\section{Related Work}\label{sec:relatedwork}
\subsection{Machine Learning-Based Vulnerability Detection}

DLVD approaches could be categorized into two families based on the way of feature extraction: token-based or graph-based approaches.
In the token-based approach~\cite{fu2022linevul,li2018vuldeepecker,li2021sysevr,rahman2024towards}, code is considered as a sequence of tokens and is represented as a vector via text embedding techniques (e.g., GloVe~\cite{pennington2014glove} and CodeBERT~\cite{feng2020codebert}).
For instance, LineVul~\cite{fu2022linevul} leverages CodeBERT~\cite{feng2020codebert} to embed the whole sequence of tokens in a function for vulnerability detection. Instead of considering the whole code, approaches such as VulDeePecker~\cite{li2018vuldeepecker} and SySeVR~\cite{li2021sysevr}, extract slices from points of interest in code (e.g., API calls, array indexing, pointer usage, etc.) and use them for vulnerability detection since they assume that different lines of code are not equivalently important for vulnerability detection.
Another family of approaches~\cite{chakraborty2021deep,zhou2019devign,li2021vulnerability, hin2022linevd} consider code as graphs and incorporate the information in different syntactic and semantic dependencies using graph neural network (GNN)~\cite{wu2020comprehensive} when generating the representation vectors. Different types of syntactic/semantic graphs, (e.g., abstract syntax tree (AST), program dependency graph (PDG), code property graph (CPG)) can be used. For example, Devign~\cite{zhou2019devign} and Reveal~\cite{chakraborty2021deep} leverage code property graph (CPG)~\cite{yamaguchi2014modeling} to build their graph-based vulnerability detection model. IVDetect~\cite{li2021vulnerability} and LineVD~\cite{hin2022linevd} consider the vulnerable statements and capture their surrounding contexts via a program dependency graph.
Both existing graph-based and text-based models follow the one-for-all design, in which only one final model is trained to handle all types of vulnerabilities. To address the limitation of previous existing approaches, we propose \ourTool to leverage the MoE framework to assign appropriate experts for handling the vulnerabilities of specific CWE types.

\subsection{Mixture-of-Experts (MoE) in Software Engineering Tasks}
The Mixture-of-Experts (MoE) is relatively new in the software engineering domain and we only found it was leveraged in defect prediction~\cite{omer2023me,shankar2023implicit}. Omer et al. presented a Mixture-of-Experts (MoE)-based approach named ME-SFP~\cite{omer2023me} to perform software defect prediction. ME-SFP uses the experts trained using f decision trees and multilayer perceptions and a Gaussian mixture model as a gating function to select appropriate experts. Aditya and Santosh~\cite{shankar2023implicit} explored two variations of the MoE method, a mixture of implicit experts (MIoE) and a mixture of explicit experts (MEoE) for the defect prediction task. The MIoE method randomly partitions the input data into a number of sub-spaces using an employed error function. The local experts become specialized in each sub-space. These specialized experts are further used for the final prediction. The MEoE method explicitly partitioned the input data into a number of sub-spaces using a clustering method before the training process. Each expert is assigned one of these sub-spaces. The local experts become specialized in these sub-spaces, which are subsequently used for the final prediction. In this study, we follow the MEoE method, in which we partition the input space based on CWE types, which is more suitable for the nature of the task.

\section{Conclusion}\label{sec:conclusion}
In this work, we present \ourTool, the first approach in vulnerability detection using the Mixture-of-Experts (MoE) framework.
\ourTool mitigates the limitations of traditional ``one-for-all'' design by leveraging specialized experts for different CWE types.
\ourTool achieved the highest F1-score of 0.44 on the BigVul vulnerability detection dataset, surpassing the best existing approach by 12.8\%. 
In handling long-tailed distributions, \ourTool maintains strong performance across both common and rare CWE types. Benefiting from its ability to learn various vulnerability patterns individually, \ourTool improved the performance on the tail group by 31.2\% in terms of recall while maintaining similar precision. 
As the first application of the MoE framework in vulnerability detection, our work suggests the possibility of leveraging specialized models and indicates potential for more advanced MoE architectures to further enhance detection capabilities.
\section*{Data Availability}
We make all the datasets, results and the code used in this study openly available in our replication package~\cite{MoEreplication}.
\bibliographystyle{ACM-Reference-Format}
\bibliography{sample-base}


\begin{thebibliography}{50}


\ifx \showCODEN    \undefined \def \showCODEN     #1{\unskip}     \fi
\ifx \showISBNx    \undefined \def \showISBNx     #1{\unskip}     \fi
\ifx \showISBNxiii \undefined \def \showISBNxiii  #1{\unskip}     \fi
\ifx \showISSN     \undefined \def \showISSN      #1{\unskip}     \fi
\ifx \showLCCN     \undefined \def \showLCCN      #1{\unskip}     \fi
\ifx \shownote     \undefined \def \shownote      #1{#1}          \fi
\ifx \showarticletitle \undefined \def \showarticletitle #1{#1}   \fi
\ifx \showURL      \undefined \def \showURL       {\relax}        \fi
\providecommand\bibfield[2]{#2}
\providecommand\bibinfo[2]{#2}
\providecommand\natexlab[1]{#1}
\providecommand\showeprint[2][]{arXiv:#2}

\bibitem[MoE(2024)]%
        {MoEreplication}
 \bibinfo{year}{2024}\natexlab{}.
\newblock \bibinfo{booktitle}{\emph{MoEVD replication pakage}}. \bibinfo{publisher}{Zenodo}.
\newblock
\href{https://doi.org/10.5281/zenodo.11661787}{doi:\nolinkurl{10.5281/zenodo.11661787}}


\bibitem[Aota et~al\mbox{.}(2020)]%
        {aota2020automation}
\bibfield{author}{\bibinfo{person}{Masaki Aota}, \bibinfo{person}{Hideaki Kanehara}, \bibinfo{person}{Masaki Kubo}, \bibinfo{person}{Noboru Murata}, \bibinfo{person}{Bo Sun}, {and} \bibinfo{person}{Takeshi Takahashi}.} \bibinfo{year}{2020}\natexlab{}.
\newblock \showarticletitle{Automation of vulnerability classification from its description using machine learning}. In \bibinfo{booktitle}{\emph{2020 IEEE Symposium on Computers and Communications (ISCC)}}. IEEE, \bibinfo{pages}{1--7}.
\newblock


\bibitem[Bent{\'e}jac et~al\mbox{.}(2021)]%
        {bentejac2021comparative}
\bibfield{author}{\bibinfo{person}{Candice Bent{\'e}jac}, \bibinfo{person}{Anna Cs{\"o}rg{\H{o}}}, {and} \bibinfo{person}{Gonzalo Mart{\'\i}nez-Mu{\~n}oz}.} \bibinfo{year}{2021}\natexlab{}.
\newblock \showarticletitle{A comparative analysis of gradient boosting algorithms}.
\newblock \bibinfo{journal}{\emph{Artificial Intelligence Review}}  \bibinfo{volume}{54} (\bibinfo{year}{2021}), \bibinfo{pages}{1937--1967}.
\newblock


\bibitem[Biau and Scornet(2016)]%
        {biau2016random}
\bibfield{author}{\bibinfo{person}{G{\'e}rard Biau} {and} \bibinfo{person}{Erwan Scornet}.} \bibinfo{year}{2016}\natexlab{}.
\newblock \showarticletitle{A random forest guided tour}.
\newblock \bibinfo{journal}{\emph{Test}}  \bibinfo{volume}{25} (\bibinfo{year}{2016}), \bibinfo{pages}{197--227}.
\newblock


\bibitem[Chakraborty et~al\mbox{.}(2021)]%
        {chakraborty2021deep}
\bibfield{author}{\bibinfo{person}{Saikat Chakraborty}, \bibinfo{person}{Rahul Krishna}, \bibinfo{person}{Yangruibo Ding}, {and} \bibinfo{person}{Baishakhi Ray}.} \bibinfo{year}{2021}\natexlab{}.
\newblock \showarticletitle{Deep learning based vulnerability detection: Are we there yet?}
\newblock \bibinfo{journal}{\emph{IEEE Transactions on Software Engineering}} \bibinfo{volume}{48}, \bibinfo{number}{9} (\bibinfo{year}{2021}), \bibinfo{pages}{3280--3296}.
\newblock


\bibitem[Chen and Guestrin(2016)]%
        {chen2016xgboost}
\bibfield{author}{\bibinfo{person}{Tianqi Chen} {and} \bibinfo{person}{Carlos Guestrin}.} \bibinfo{year}{2016}\natexlab{}.
\newblock \showarticletitle{Xgboost: A scalable tree boosting system}. In \bibinfo{booktitle}{\emph{Proceedings of the 22nd acm sigkdd international conference on knowledge discovery and data mining}}. \bibinfo{pages}{785--794}.
\newblock


\bibitem[Christey et~al\mbox{.}(2013)]%
        {CWE}
\bibfield{author}{\bibinfo{person}{Steve Christey}, \bibinfo{person}{J Kenderdine}, \bibinfo{person}{J Mazella}, {and} \bibinfo{person}{B Miles}.} \bibinfo{year}{2013}\natexlab{}.
\newblock \showarticletitle{Common weakness enumeration}.
\newblock \bibinfo{journal}{\emph{Mitre Corporation}} (\bibinfo{year}{2013}).
\newblock


\bibitem[Devlin et~al\mbox{.}(2019)]%
        {devlin2018bert}
\bibfield{author}{\bibinfo{person}{Jacob Devlin}, \bibinfo{person}{Ming-Wei Chang}, \bibinfo{person}{Kenton Lee}, {and} \bibinfo{person}{Kristina Toutanova}.} \bibinfo{year}{2019}\natexlab{}.
\newblock \showarticletitle{{BERT}: Pre-training of Deep Bidirectional Transformers for Language Understanding}. In \bibinfo{booktitle}{\emph{Proceedings of the 17th 2019 Conference of the North {A}merican Chapter of the Association for Computational Linguistics: Human Language Technologies (HLT-NAACL)}}. \bibinfo{pages}{4171--4186}.
\newblock


\bibitem[Erickson et~al\mbox{.}(2020a)]%
        {erickson2020autogluon}
\bibfield{author}{\bibinfo{person}{Nick Erickson}, \bibinfo{person}{Jonas Mueller}, \bibinfo{person}{Alexander Shirkov}, \bibinfo{person}{Hang Zhang}, \bibinfo{person}{Pedro Larroy}, \bibinfo{person}{Mu Li}, {and} \bibinfo{person}{Alexander Smola}.} \bibinfo{year}{2020}\natexlab{a}.
\newblock \showarticletitle{Autogluon-tabular: Robust and accurate automl for structured data}.
\newblock \bibinfo{journal}{\emph{arXiv preprint arXiv:2003.06505}} (\bibinfo{year}{2020}).
\newblock


\bibitem[Erickson et~al\mbox{.}(2020b)]%
        {agtabular}
\bibfield{author}{\bibinfo{person}{Nick Erickson}, \bibinfo{person}{Jonas Mueller}, \bibinfo{person}{Alexander Shirkov}, \bibinfo{person}{Hang Zhang}, \bibinfo{person}{Pedro Larroy}, \bibinfo{person}{Mu Li}, {and} \bibinfo{person}{Alexander Smola}.} \bibinfo{year}{2020}\natexlab{b}.
\newblock \showarticletitle{AutoGluon-Tabular: Robust and Accurate AutoML for Structured Data}.
\newblock \bibinfo{journal}{\emph{arXiv preprint arXiv:2003.06505}} (\bibinfo{year}{2020}).
\newblock


\bibitem[Fan et~al\mbox{.}(2020)]%
        {fan2020ac}
\bibfield{author}{\bibinfo{person}{Jiahao Fan}, \bibinfo{person}{Yi Li}, \bibinfo{person}{Shaohua Wang}, {and} \bibinfo{person}{Tien~N Nguyen}.} \bibinfo{year}{2020}\natexlab{}.
\newblock \showarticletitle{AC/C++ code vulnerability dataset with code changes and CVE summaries}. In \bibinfo{booktitle}{\emph{Proceedings of the 17th International Conference on Mining Software Repositories}}. \bibinfo{pages}{508--512}.
\newblock


\bibitem[Fedus et~al\mbox{.}(2022)]%
        {fedus2022switch}
\bibfield{author}{\bibinfo{person}{William Fedus}, \bibinfo{person}{Barret Zoph}, {and} \bibinfo{person}{Noam Shazeer}.} \bibinfo{year}{2022}\natexlab{}.
\newblock \showarticletitle{Switch transformers: Scaling to trillion parameter models with simple and efficient sparsity}.
\newblock \bibinfo{journal}{\emph{Journal of Machine Learning Research}} \bibinfo{volume}{23}, \bibinfo{number}{120} (\bibinfo{year}{2022}), \bibinfo{pages}{1--39}.
\newblock


\bibitem[Feng et~al\mbox{.}(2020)]%
        {feng2020codebert}
\bibfield{author}{\bibinfo{person}{Zhangyin Feng}, \bibinfo{person}{Daya Guo}, \bibinfo{person}{Duyu Tang}, \bibinfo{person}{Nan Duan}, \bibinfo{person}{Xiaocheng Feng}, \bibinfo{person}{Ming Gong}, \bibinfo{person}{Linjun Shou}, \bibinfo{person}{Bing Qin}, \bibinfo{person}{Ting Liu}, \bibinfo{person}{Daxin Jiang}, {and} \bibinfo{person}{Ming Zhou}.} \bibinfo{year}{2020}\natexlab{}.
\newblock \showarticletitle{CodeBERT: A Pre-Trained Model for Programming and Natural Languages}. In \bibinfo{booktitle}{\emph{Findings of EMNLP}}.
\newblock


\bibitem[focal loss(2024)]%
        {focalloss}
\bibfield{author}{\bibinfo{person}{focal loss}.} \bibinfo{year}{2024}\natexlab{}.
\newblock \bibinfo{title}{Focal Loss Documentation}.
\newblock \bibinfo{howpublished}{\url{https://focal-loss.readthedocs.io/en/latest/generated/focal_loss.binary_focal_loss.html}}.
\newblock
\newblock
\shownote{Accessed: 2024-06-08}.


\bibitem[Fu et~al\mbox{.}(2023)]%
        {fu2023vulexplainer}
\bibfield{author}{\bibinfo{person}{Michael Fu}, \bibinfo{person}{Van Nguyen}, \bibinfo{person}{Chakkrit~Kla Tantithamthavorn}, \bibinfo{person}{Trung Le}, {and} \bibinfo{person}{Dinh Phung}.} \bibinfo{year}{2023}\natexlab{}.
\newblock \showarticletitle{Vulexplainer: A transformer-based hierarchical distillation for explaining vulnerability types}.
\newblock \bibinfo{journal}{\emph{IEEE Transactions on Software Engineering}} (\bibinfo{year}{2023}).
\newblock


\bibitem[Fu and Tantithamthavorn(2022)]%
        {fu2022linevul}
\bibfield{author}{\bibinfo{person}{Michael Fu} {and} \bibinfo{person}{Chakkrit Tantithamthavorn}.} \bibinfo{year}{2022}\natexlab{}.
\newblock \showarticletitle{Linevul: A transformer-based line-level vulnerability prediction}. In \bibinfo{booktitle}{\emph{Proceedings of the 19th International Conference on Mining Software Repositories}}. \bibinfo{pages}{608--620}.
\newblock


\bibitem[Guo et~al\mbox{.}(2022)]%
        {guo2022unixcoder}
\bibfield{author}{\bibinfo{person}{Daya Guo}, \bibinfo{person}{Shuai Lu}, \bibinfo{person}{Nan Duan}, \bibinfo{person}{Yanlin Wang}, \bibinfo{person}{Ming Zhou}, {and} \bibinfo{person}{Jian Yin}.} \bibinfo{year}{2022}\natexlab{}.
\newblock \showarticletitle{Unixcoder: Unified cross-modal pre-training for code representation}.
\newblock \bibinfo{journal}{\emph{arXiv preprint arXiv:2203.03850}} (\bibinfo{year}{2022}).
\newblock


\bibitem[Guo et~al\mbox{.}(2020)]%
        {guo2020graphcodebert}
\bibfield{author}{\bibinfo{person}{Daya Guo}, \bibinfo{person}{Shuo Ren}, \bibinfo{person}{Shuai Lu}, \bibinfo{person}{Zhangyin Feng}, \bibinfo{person}{Duyu Tang}, \bibinfo{person}{Shujie Liu}, \bibinfo{person}{Long Zhou}, \bibinfo{person}{Nan Duan}, \bibinfo{person}{Alexey Svyatkovskiy}, \bibinfo{person}{Shengyu Fu}, {et~al\mbox{.}}} \bibinfo{year}{2020}\natexlab{}.
\newblock \showarticletitle{Graphcodebert: Pre-training code representations with data flow}.
\newblock \bibinfo{journal}{\emph{arXiv preprint arXiv:2009.08366}} (\bibinfo{year}{2020}).
\newblock


\bibitem[He and Garcia(2009)]%
        {he2009learning}
\bibfield{author}{\bibinfo{person}{Haibo He} {and} \bibinfo{person}{Edwardo~A Garcia}.} \bibinfo{year}{2009}\natexlab{}.
\newblock \showarticletitle{Learning from imbalanced data}.
\newblock \bibinfo{journal}{\emph{IEEE Transactions on knowledge and data engineering}} \bibinfo{volume}{21}, \bibinfo{number}{9} (\bibinfo{year}{2009}), \bibinfo{pages}{1263--1284}.
\newblock


\bibitem[Hin et~al\mbox{.}(2022)]%
        {hin2022linevd}
\bibfield{author}{\bibinfo{person}{David Hin}, \bibinfo{person}{Andrey Kan}, \bibinfo{person}{Huaming Chen}, {and} \bibinfo{person}{M~Ali Babar}.} \bibinfo{year}{2022}\natexlab{}.
\newblock \showarticletitle{Linevd: Statement-level vulnerability detection using graph neural networks}. In \bibinfo{booktitle}{\emph{Proceedings of the 19th international conference on mining software repositories}}. \bibinfo{pages}{596--607}.
\newblock


\bibitem[Jacobs et~al\mbox{.}(1991)]%
        {jacobs1991textordfeminineadaptive}
\bibfield{author}{\bibinfo{person}{RA Jacobs}, \bibinfo{person}{MI Jordan}, \bibinfo{person}{SJ Nowlan}, {and} \bibinfo{person}{GE Hinton}.} \bibinfo{year}{1991}\natexlab{}.
\newblock \showarticletitle{{\textordfeminine}Adaptive Mixtures of Local Experts, {\textordmasculine} Neural Computation, vol. 3}.
\newblock  (\bibinfo{year}{1991}).
\newblock


\bibitem[Ke et~al\mbox{.}(2017)]%
        {ke2017lightgbm}
\bibfield{author}{\bibinfo{person}{Guolin Ke}, \bibinfo{person}{Qi Meng}, \bibinfo{person}{Thomas Finley}, \bibinfo{person}{Taifeng Wang}, \bibinfo{person}{Wei Chen}, \bibinfo{person}{Weidong Ma}, \bibinfo{person}{Qiwei Ye}, {and} \bibinfo{person}{Tie-Yan Liu}.} \bibinfo{year}{2017}\natexlab{}.
\newblock \showarticletitle{Lightgbm: A highly efficient gradient boosting decision tree}.
\newblock \bibinfo{journal}{\emph{Advances in neural information processing systems}}  \bibinfo{volume}{30} (\bibinfo{year}{2017}).
\newblock


\bibitem[Lepikhin et~al\mbox{.}(2020)]%
        {lepikhin2020gshard}
\bibfield{author}{\bibinfo{person}{Dmitry Lepikhin}, \bibinfo{person}{HyoukJoong Lee}, \bibinfo{person}{Yuanzhong Xu}, \bibinfo{person}{Dehao Chen}, \bibinfo{person}{Orhan Firat}, \bibinfo{person}{Yanping Huang}, \bibinfo{person}{Maxim Krikun}, \bibinfo{person}{Noam Shazeer}, {and} \bibinfo{person}{Zhifeng Chen}.} \bibinfo{year}{2020}\natexlab{}.
\newblock \showarticletitle{Gshard: Scaling giant models with conditional computation and automatic sharding}.
\newblock \bibinfo{journal}{\emph{arXiv preprint arXiv:2006.16668}} (\bibinfo{year}{2020}).
\newblock


\bibitem[Li et~al\mbox{.}(2021a)]%
        {li2021vulnerability}
\bibfield{author}{\bibinfo{person}{Yi Li}, \bibinfo{person}{Shaohua Wang}, {and} \bibinfo{person}{Tien~N Nguyen}.} \bibinfo{year}{2021}\natexlab{a}.
\newblock \showarticletitle{Vulnerability detection with fine-grained interpretations}. In \bibinfo{booktitle}{\emph{Proceedings of the 29th ACM Joint Meeting on European Software Engineering Conference and Symposium on the Foundations of Software Engineering}}. \bibinfo{pages}{292--303}.
\newblock


\bibitem[Li et~al\mbox{.}(2021b)]%
        {li2021sysevr}
\bibfield{author}{\bibinfo{person}{Zhen Li}, \bibinfo{person}{Deqing Zou}, \bibinfo{person}{Shouhuai Xu}, \bibinfo{person}{Hai Jin}, \bibinfo{person}{Yawei Zhu}, {and} \bibinfo{person}{Zhaoxuan Chen}.} \bibinfo{year}{2021}\natexlab{b}.
\newblock \showarticletitle{Sysevr: A framework for using deep learning to detect software vulnerabilities}.
\newblock \bibinfo{journal}{\emph{IEEE Transactions on Dependable and Secure Computing}} (\bibinfo{year}{2021}).
\newblock


\bibitem[Li et~al\mbox{.}(2018)]%
        {li2018vuldeepecker}
\bibfield{author}{\bibinfo{person}{Zhen Li}, \bibinfo{person}{Deqing Zou}, \bibinfo{person}{Shouhuai Xu}, \bibinfo{person}{Xinyu Ou}, \bibinfo{person}{Hai Jin}, \bibinfo{person}{Sujuan Wang}, \bibinfo{person}{Zhijun Deng}, {and} \bibinfo{person}{Yuyi Zhong}.} \bibinfo{year}{2018}\natexlab{}.
\newblock \showarticletitle{VulDeePecker: {A} Deep Learning-Based System for Vulnerability Detection}. In \bibinfo{booktitle}{\emph{25th Annual Network and Distributed System Security Symposium, {NDSS} 2018, San Diego, California, USA, February 18-21, 2018}}.
\newblock


\bibitem[Lin et~al\mbox{.}(2017)]%
        {lin2017focal}
\bibfield{author}{\bibinfo{person}{Tsung-Yi Lin}, \bibinfo{person}{Priya Goyal}, \bibinfo{person}{Ross Girshick}, \bibinfo{person}{Kaiming He}, {and} \bibinfo{person}{Piotr Doll{\'a}r}.} \bibinfo{year}{2017}\natexlab{}.
\newblock \showarticletitle{Focal loss for dense object detection}. In \bibinfo{booktitle}{\emph{Proceedings of the IEEE international conference on computer vision}}. \bibinfo{pages}{2980--2988}.
\newblock


\bibitem[Loshchilov and Hutter(2017)]%
        {loshchilov2017decoupled}
\bibfield{author}{\bibinfo{person}{Ilya Loshchilov} {and} \bibinfo{person}{Frank Hutter}.} \bibinfo{year}{2017}\natexlab{}.
\newblock \showarticletitle{Decoupled weight decay regularization}.
\newblock \bibinfo{journal}{\emph{arXiv preprint arXiv:1711.05101}} (\bibinfo{year}{2017}).
\newblock


\bibitem[MITRE(2024)]%
        {cwe.top.25}
\bibfield{author}{\bibinfo{person}{MITRE}.} \bibinfo{year}{2024}\natexlab{}.
\newblock \bibinfo{title}{CWE Top 25 Most Dangerous Software Weaknesses}.
\newblock
\urldef\tempurl%
\url{https://cwe.mitre.org/top25/}
\showURL{%
\tempurl}


\bibitem[Nguyen et~al\mbox{.}(2022)]%
        {nguyen2022regvd}
\bibfield{author}{\bibinfo{person}{Van-Anh Nguyen}, \bibinfo{person}{Dai~Quoc Nguyen}, \bibinfo{person}{Van Nguyen}, \bibinfo{person}{Trung Le}, \bibinfo{person}{Quan~Hung Tran}, {and} \bibinfo{person}{Dinh Phung}.} \bibinfo{year}{2022}\natexlab{}.
\newblock \showarticletitle{Regvd: Revisiting graph neural networks for vulnerability detection}. In \bibinfo{booktitle}{\emph{Proceedings of the ACM/IEEE 44th International Conference on Software Engineering: Companion Proceedings}}. \bibinfo{pages}{178--182}.
\newblock


\bibitem[Omer et~al\mbox{.}(2023)]%
        {omer2023me}
\bibfield{author}{\bibinfo{person}{Aman Omer}, \bibinfo{person}{Santosh~Singh Rathore}, {and} \bibinfo{person}{Sandeep Kumar}.} \bibinfo{year}{2023}\natexlab{}.
\newblock \showarticletitle{ME-SFP: A Mixture-of-Experts-Based Approach for Software Fault Prediction}.
\newblock \bibinfo{journal}{\emph{IEEE Transactions on Reliability}} (\bibinfo{year}{2023}).
\newblock


\bibitem[Paszke et~al\mbox{.}(2019)]%
        {paszke2019pytorch}
\bibfield{author}{\bibinfo{person}{Adam Paszke}, \bibinfo{person}{Sam Gross}, \bibinfo{person}{Francisco Massa}, \bibinfo{person}{Adam Lerer}, \bibinfo{person}{James Bradbury}, \bibinfo{person}{Gregory Chanan}, \bibinfo{person}{Trevor Killeen}, \bibinfo{person}{Zeming Lin}, \bibinfo{person}{Natalia Gimelshein}, \bibinfo{person}{Luca Antiga}, {et~al\mbox{.}}} \bibinfo{year}{2019}\natexlab{}.
\newblock \showarticletitle{Pytorch: An imperative style, high-performance deep learning library}.
\newblock \bibinfo{journal}{\emph{Advances in neural information processing systems}}  \bibinfo{volume}{32} (\bibinfo{year}{2019}).
\newblock


\bibitem[Pennington et~al\mbox{.}(2014)]%
        {pennington2014glove}
\bibfield{author}{\bibinfo{person}{Jeffrey Pennington}, \bibinfo{person}{Richard Socher}, {and} \bibinfo{person}{Christopher~D Manning}.} \bibinfo{year}{2014}\natexlab{}.
\newblock \showarticletitle{Glove: Global vectors for word representation}. In \bibinfo{booktitle}{\emph{Proceedings of the 2014 conference on empirical methods in natural language processing (EMNLP)}}. \bibinfo{pages}{1532--1543}.
\newblock


\bibitem[Prokhorenkova et~al\mbox{.}(2018)]%
        {prokhorenkova2018catboost}
\bibfield{author}{\bibinfo{person}{Liudmila Prokhorenkova}, \bibinfo{person}{Gleb Gusev}, \bibinfo{person}{Aleksandr Vorobev}, \bibinfo{person}{Anna~Veronika Dorogush}, {and} \bibinfo{person}{Andrey Gulin}.} \bibinfo{year}{2018}\natexlab{}.
\newblock \showarticletitle{CatBoost: unbiased boosting with categorical features}.
\newblock \bibinfo{journal}{\emph{Advances in neural information processing systems}}  \bibinfo{volume}{31} (\bibinfo{year}{2018}).
\newblock


\bibitem[Rahman et~al\mbox{.}(2024)]%
        {rahman2024towards}
\bibfield{author}{\bibinfo{person}{Md~Mahbubur Rahman}, \bibinfo{person}{Ira Ceka}, \bibinfo{person}{Chengzhi Mao}, \bibinfo{person}{Saikat Chakraborty}, \bibinfo{person}{Baishakhi Ray}, {and} \bibinfo{person}{Wei Le}.} \bibinfo{year}{2024}\natexlab{}.
\newblock \showarticletitle{Towards causal deep learning for vulnerability detection}. In \bibinfo{booktitle}{\emph{Proceedings of the IEEE/ACM 46th International Conference on Software Engineering}}. \bibinfo{pages}{1--11}.
\newblock


\bibitem[Scandariato et~al\mbox{.}(2014)]%
        {scandariato2014predicting}
\bibfield{author}{\bibinfo{person}{Riccardo Scandariato}, \bibinfo{person}{James Walden}, \bibinfo{person}{Aram Hovsepyan}, {and} \bibinfo{person}{Wouter Joosen}.} \bibinfo{year}{2014}\natexlab{}.
\newblock \showarticletitle{Predicting vulnerable software components via text mining}.
\newblock \bibinfo{journal}{\emph{IEEE Transactions on Software Engineering}} \bibinfo{volume}{40}, \bibinfo{number}{10} (\bibinfo{year}{2014}), \bibinfo{pages}{993--1006}.
\newblock


\bibitem[Scarselli et~al\mbox{.}(2008)]%
        {scarselli2008graph}
\bibfield{author}{\bibinfo{person}{Franco Scarselli}, \bibinfo{person}{Marco Gori}, \bibinfo{person}{Ah~Chung Tsoi}, \bibinfo{person}{Markus Hagenbuchner}, {and} \bibinfo{person}{Gabriele Monfardini}.} \bibinfo{year}{2008}\natexlab{}.
\newblock \showarticletitle{The graph neural network model}.
\newblock \bibinfo{journal}{\emph{IEEE transactions on neural networks}} \bibinfo{volume}{20}, \bibinfo{number}{1} (\bibinfo{year}{2008}), \bibinfo{pages}{61--80}.
\newblock


\bibitem[Shankar~Mishra and Singh~Rathore(2023)]%
        {shankar2023implicit}
\bibfield{author}{\bibinfo{person}{Aditya Shankar~Mishra} {and} \bibinfo{person}{Santosh Singh~Rathore}.} \bibinfo{year}{2023}\natexlab{}.
\newblock \showarticletitle{Implicit and explicit mixture of experts models for software defect prediction}.
\newblock \bibinfo{journal}{\emph{Software Quality Journal}} \bibinfo{volume}{31}, \bibinfo{number}{4} (\bibinfo{year}{2023}), \bibinfo{pages}{1331--1368}.
\newblock


\bibitem[Shazeer et~al\mbox{.}(2017)]%
        {shazeer2017outrageously}
\bibfield{author}{\bibinfo{person}{Noam Shazeer}, \bibinfo{person}{Azalia Mirhoseini}, \bibinfo{person}{Krzysztof Maziarz}, \bibinfo{person}{Andy Davis}, \bibinfo{person}{Quoc Le}, \bibinfo{person}{Geoffrey Hinton}, {and} \bibinfo{person}{Jeff Dean}.} \bibinfo{year}{2017}\natexlab{}.
\newblock \showarticletitle{Outrageously large neural networks: The sparsely-gated mixture-of-experts layer}.
\newblock \bibinfo{journal}{\emph{arXiv preprint arXiv:1701.06538}} (\bibinfo{year}{2017}).
\newblock


\bibitem[Szczepanek(2022)]%
        {szczepanek2022daily}
\bibfield{author}{\bibinfo{person}{Robert Szczepanek}.} \bibinfo{year}{2022}\natexlab{}.
\newblock \showarticletitle{Daily streamflow forecasting in mountainous catchment using XGBoost, LightGBM and CatBoost}.
\newblock \bibinfo{journal}{\emph{Hydrology}} \bibinfo{volume}{9}, \bibinfo{number}{12} (\bibinfo{year}{2022}), \bibinfo{pages}{226}.
\newblock


\bibitem[Wan et~al\mbox{.}(2024)]%
        {wan2024bridging}
\bibfield{author}{\bibinfo{person}{Shengye Wan}, \bibinfo{person}{Joshua Saxe}, \bibinfo{person}{Craig Gomes}, \bibinfo{person}{Sahana Chennabasappa}, \bibinfo{person}{Avilash Rath}, \bibinfo{person}{Kun Sun}, {and} \bibinfo{person}{Xinda Wang}.} \bibinfo{year}{2024}\natexlab{}.
\newblock \showarticletitle{Bridging the Gap: A Study of AI-based Vulnerability Management between Industry and Academia}.
\newblock \bibinfo{journal}{\emph{arXiv preprint arXiv:2405.02435}} (\bibinfo{year}{2024}).
\newblock


\bibitem[Wen et~al\mbox{.}(2024)]%
        {wen2024livable}
\bibfield{author}{\bibinfo{person}{Xin-Cheng Wen}, \bibinfo{person}{Cuiyun Gao}, \bibinfo{person}{Feng Luo}, \bibinfo{person}{Haoyu Wang}, \bibinfo{person}{Ge Li}, {and} \bibinfo{person}{Qing Liao}.} \bibinfo{year}{2024}\natexlab{}.
\newblock \showarticletitle{LIVABLE: exploring long-tailed classification of software vulnerability types}.
\newblock \bibinfo{journal}{\emph{IEEE Transactions on Software Engineering}} (\bibinfo{year}{2024}).
\newblock


\bibitem[Wolf et~al\mbox{.}(2019)]%
        {wolf2019huggingface}
\bibfield{author}{\bibinfo{person}{Thomas Wolf}, \bibinfo{person}{Lysandre Debut}, \bibinfo{person}{Victor Sanh}, \bibinfo{person}{Julien Chaumond}, \bibinfo{person}{Clement Delangue}, \bibinfo{person}{Anthony Moi}, \bibinfo{person}{Pierric Cistac}, \bibinfo{person}{Tim Rault}, \bibinfo{person}{R{\'e}mi Louf}, \bibinfo{person}{Morgan Funtowicz}, {et~al\mbox{.}}} \bibinfo{year}{2019}\natexlab{}.
\newblock \showarticletitle{Huggingface's transformers: State-of-the-art natural language processing}.
\newblock \bibinfo{journal}{\emph{arXiv preprint arXiv:1910.03771}} (\bibinfo{year}{2019}).
\newblock


\bibitem[Wu et~al\mbox{.}(2020)]%
        {wu2020comprehensive}
\bibfield{author}{\bibinfo{person}{Zonghan Wu}, \bibinfo{person}{Shirui Pan}, \bibinfo{person}{Fengwen Chen}, \bibinfo{person}{Guodong Long}, \bibinfo{person}{Chengqi Zhang}, {and} \bibinfo{person}{S~Yu Philip}.} \bibinfo{year}{2020}\natexlab{}.
\newblock \showarticletitle{A comprehensive survey on graph neural networks}.
\newblock \bibinfo{journal}{\emph{IEEE transactions on neural networks and learning systems}} \bibinfo{volume}{32}, \bibinfo{number}{1} (\bibinfo{year}{2020}), \bibinfo{pages}{4--24}.
\newblock


\bibitem[Yamaguchi et~al\mbox{.}(2014)]%
        {yamaguchi2014modeling}
\bibfield{author}{\bibinfo{person}{Fabian Yamaguchi}, \bibinfo{person}{Nico Golde}, \bibinfo{person}{Daniel Arp}, {and} \bibinfo{person}{Konrad Rieck}.} \bibinfo{year}{2014}\natexlab{}.
\newblock \showarticletitle{Modeling and discovering vulnerabilities with code property graphs}. In \bibinfo{booktitle}{\emph{2014 IEEE Symposium on Security and Privacy}}. IEEE, \bibinfo{pages}{590--604}.
\newblock


\bibitem[Yang et~al\mbox{.}(2023)]%
        {yang2023does}
\bibfield{author}{\bibinfo{person}{Xu Yang}, \bibinfo{person}{Shaowei Wang}, \bibinfo{person}{Yi Li}, {and} \bibinfo{person}{Shaohua Wang}.} \bibinfo{year}{2023}\natexlab{}.
\newblock \showarticletitle{Does data sampling improve deep learning-based vulnerability detection? Yeas! and Nays!}. In \bibinfo{booktitle}{\emph{Proceedings of the 45th IEEE/ACM International Conference on Software Engineering (ICSE)}}. IEEE, \bibinfo{pages}{2287--2298}.
\newblock


\bibitem[Yuksel et~al\mbox{.}(2012)]%
        {yuksel2012twenty}
\bibfield{author}{\bibinfo{person}{Seniha~Esen Yuksel}, \bibinfo{person}{Joseph~N Wilson}, {and} \bibinfo{person}{Paul~D Gader}.} \bibinfo{year}{2012}\natexlab{}.
\newblock \showarticletitle{Twenty years of mixture of experts}.
\newblock \bibinfo{journal}{\emph{IEEE transactions on neural networks and learning systems}} \bibinfo{volume}{23}, \bibinfo{number}{8} (\bibinfo{year}{2012}), \bibinfo{pages}{1177--1193}.
\newblock


\bibitem[Zhang et~al\mbox{.}(2020)]%
        {zhang2020learning}
\bibfield{author}{\bibinfo{person}{Yakun Zhang}, \bibinfo{person}{Wensheng Dou}, \bibinfo{person}{Jiaxin Zhu}, \bibinfo{person}{Liang Xu}, \bibinfo{person}{Zhiyong Zhou}, \bibinfo{person}{Jun Wei}, \bibinfo{person}{Dan Ye}, {and} \bibinfo{person}{Bo Yang}.} \bibinfo{year}{2020}\natexlab{}.
\newblock \showarticletitle{Learning to detect table clones in spreadsheets}. In \bibinfo{booktitle}{\emph{Proceedings of the 29th ACM SIGSOFT International Symposium on Software Testing and Analysis}}. \bibinfo{pages}{528--540}.
\newblock


\bibitem[Zhou et~al\mbox{.}(2023)]%
        {zhou2023devil}
\bibfield{author}{\bibinfo{person}{Xin Zhou}, \bibinfo{person}{Kisub Kim}, \bibinfo{person}{Bowen Xu}, \bibinfo{person}{Jiakun Liu}, \bibinfo{person}{DongGyun Han}, {and} \bibinfo{person}{David Lo}.} \bibinfo{year}{2023}\natexlab{}.
\newblock \showarticletitle{The devil is in the tails: How long-tailed code distributions impact large language models}. In \bibinfo{booktitle}{\emph{2023 38th IEEE/ACM International Conference on Automated Software Engineering (ASE)}}. IEEE, \bibinfo{pages}{40--52}.
\newblock


\bibitem[Zhou et~al\mbox{.}(2019)]%
        {zhou2019devign}
\bibfield{author}{\bibinfo{person}{Yaqin Zhou}, \bibinfo{person}{Shangqing Liu}, \bibinfo{person}{Jingkai Siow}, \bibinfo{person}{Xiaoning Du}, {and} \bibinfo{person}{Yang Liu}.} \bibinfo{year}{2019}\natexlab{}.
\newblock \showarticletitle{Devign: Effective vulnerability identification by learning comprehensive program semantics via graph neural networks}.
\newblock \bibinfo{journal}{\emph{Advances in neural information processing systems}}  \bibinfo{volume}{32} (\bibinfo{year}{2019}).
\newblock


\end{thebibliography}


\end{document}